\documentclass[12pt]{iopart}

\pdfoutput=1

\usepackage{graphicx,sidecap}
\usepackage{bm}
\usepackage{braket}
\usepackage{amssymb}
\usepackage{mathrsfs}
\usepackage{amsmath}
\usepackage{xcolor}
\usepackage{hyperref}
\usepackage{enumerate}
\usepackage[toc,title]{appendix}

\hypersetup{colorlinks,bookmarksopen,bookmarksnumbered,
citecolor=red,
linkcolor=blue,
pdfstartview=green,
urlcolor=purple}

\catcode`@=11 % If we need private TeX macros
\renewcommand\tableofcontents{%
  \section*{\contentsname}%
  \@starttoc{toc}%
}
\catcode`@=12% '@' is no more a character

\def\be{\begin{equation}}
\def\ee{\end{equation}}

\def\bea{\begin{eqnarray}}
\def\eea{\end{eqnarray}}

\def\Tr{{\rm Tr}}

\usepackage{xparse}

\DeclareDocumentCommand{\TrProd}{ m O{} o O{} o O{} o O{} o }{%
\{ \Gamma_{#1}^{#2}
\IfNoValueTF{#3}{}{,\Gamma_{#3}^{#4}}
\IfNoValueTF{#5}{}{,\Gamma_{#5}^{#6}}
\IfNoValueTF{#7}{}{,\Gamma_{#7}^{#8}}
\IfNoValueTF{#9}{}{,\Gamma_{#9}}
\}
}

\DeclareDocumentCommand{\TrProdTilde}{ m O{} o O{} o O{} o O{} o }{%
\{ \widetilde\Gamma_{#1}^{#2}
\IfNoValueTF{#3}{}{,\widetilde\Gamma_{#3}^{#4}}
\IfNoValueTF{#5}{}{,\widetilde\Gamma_{#5}^{#6}}
\IfNoValueTF{#7}{}{,\widetilde\Gamma_{#7}^{#8}}
\IfNoValueTF{#9}{}{,\widetilde\Gamma_{#9}}
\}
}

\DeclareDocumentCommand{\ch}{ m o o o m o o o }{%
\begin{bmatrix}
#1 \IfNoValueTF{#2}{}{& #2}\IfNoValueTF{#3}{}{& #3}\IfNoValueTF{#4}{}{& #4} \\
#5 \IfNoValueTF{#6}{}{& #6}\IfNoValueTF{#7}{}{& #7}\IfNoValueTF{#8}{}{& #8}
\end{bmatrix}_{\tau}
}

\newcommand{\conj}[1]{#1^{\ast}}
\newcommand{\intf}[1]{\int\text{d}\conj{#1}\text{d}#1\,}
\newcommand{\df}[1]{\text{d}\conj{#1}\text{d}#1\,}
\newcommand{\iu}{\text{i}}
\makeatletter
\newcommand{\pushright}[1]{\ifmeasuring@#1\else\omit\hfill$\displaystyle#1$\fi\ignorespaces}
\newcommand{\pushleft}[1]{\ifmeasuring@#1\else\omit$\displaystyle#1$\hfill\fi\ignorespaces}
\makeatother

\begin{document}

\title[
Spin structures and entanglement of two disjoint intervals in CFT
]{
\\
Spin structures and entanglement of two disjoint intervals in conformal field theories
}

\vspace{.5cm}

\author{Andrea Coser, Erik Tonni and Pasquale Calabrese}
\address{SISSA and INFN, via Bonomea 265, 34136 Trieste, Italy. }

%\date{\today}
\vspace{.5cm}

\begin{abstract}
We reconsider the moments of the reduced density matrix of two disjoint intervals and of its partial transpose with respect to one interval for critical free fermionic lattice models. It is known that these matrices are sums of either two or four Gaussian matrices and hence their moments can be reconstructed as computable sums of products of Gaussian operators. We find that, in the scaling limit, each term in these sums is in one-to-one correspondence with the partition function of the corresponding conformal field theory on the underlying Riemann surface with a given spin structure. The analytical findings have been checked against numerical results for the Ising chain and for the XX spin chain at the critical point. 
\end{abstract}

\maketitle

\tableofcontents

\section{Introduction}

The entanglement measures for extended quantum systems have attracted a lot of interest during the last decade 
in the theoretical research of condensed matter, quantum information, quantum field theory and quantum gravity (see \cite{rev} for reviews).
Recently, interesting results have been obtained in the experimental detection of entanglement \cite{ent-experiment}.

Given a quantum system in a pure state (e.g. the ground state $|\Psi \rangle$) whose Hilbert space is spatially bipartite, i.e. $\mathcal{H} = \mathcal{H}_A \otimes \mathcal{H}_B$ at some fixed time,
in order to quantify the bipartite entanglement a crucial quantity to introduce is the reduced density matrix $\rho_A={\rm Tr}_B \rho={\rm Tr}_B |\Psi\rangle\langle \Psi|$.
In this case,  a good measure of entanglement from the quantum information perspective is the entanglement entropy
\be
\label{S_A def}
S_A = -{\rm Tr} \rho_A \log \rho_A\,,
\ee
i.e. the Von Neumann entropy for the reduced density matrix.
A very useful trick to compute the entanglement entropy is to take the replica limit $S_A =\lim_{n \to 1} S_A^{(n)}$, where $S_A^{(n)}$ are the R\'enyi entropies
\be
\label{Sndef}
S_A^{(n)} = \frac1{1-n} \, \log \Tr \rho_A^n \,,
\ee
being $\Tr \rho_A^n $ the $n$-th moment of the reduced density matrix.
When the low energy regime properties of the critical extended system is described by a $1+1$ dimensional conformal field theory (CFT) and the subsystem $A$ is a continuous interval, the entanglement entropy diverges logarithmically with the size of the subsystem and the coefficient of such divergence is proportional to the central charge $c$ of the model \cite{Holzhey,vlrk-03,cc-04,cc-rev}.

\begin{figure}
\vspace{.4cm}
\begin{center}
\includegraphics[width=0.9\textwidth]{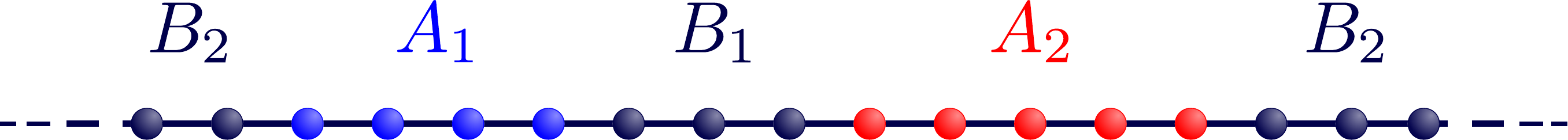}
\end{center}
\caption{
The subsystem $A = A_1 \cup A_2$ considered in this manuscript is given by two disjoint spin blocks $A_1$ and $A_2$ embedded in a spin chain of arbitrary length.
The reminder of the system is denoted by $B$ and it is also made by two disconnected pieces $B_1$ and $B_2$.
}
\label{fig:sc}
\end{figure}

In this manuscript we will be interested in the case of a subsystem $A =A_1 \cup A_2$ made by two disjoint intervals.  
The whole system is in the ground state and the Hilbert space is tripartite $\mathcal{H} = \mathcal{H}_{A_1} \otimes \mathcal{H}_{A_2} \otimes \mathcal{H}_B$.
Tracing out the degrees of freedom in $B$, we are left with the reduced density matrix $\rho_{A}$ and $S_A$ measures the entanglement between $A_1 \cup A_2$ and $B$.
On the lattice, we will consider the configuration shown in Fig.\,\ref{fig:sc}: a spin chain divided in two complementary parts $A$  and $B$, where each of them is composed by two disconnected blocks.
In the scaling limit, $A$ is given by two disjoint intervals $A_1$ and $A_2$ on the infinite line separated by the interval $B_1$, while $B_2$ has infinite length. 
A very useful quantity to  introduce in this case is the mutual information $I_{A_1, A_2} = S_{A_1} + S_{A_2} - S_{A_1 \cup A_2}$, which is UV finite in quantum field theories.

When the extended quantum system is in a mixed state, evaluating the bipartite entanglement is more complicated. 
A natural situation is the bipartite entanglement for a system in the thermal state, but an interesting setup to consider is also the one described above, where the reduced density matrix $\rho_A$ characterises a mixed state and one looks for the entanglement between $A_1$ and $A_2$.
In this manuscript we will consider only the latter case. 
A useful way to address the bipartite entanglement for $\rho_A$ is to consider its partial transpose with respect to one of the disjoint intervals.
The occurrence of negative eigenvalues in its spectrum indicates the presence of bipartite entanglement \cite{partial}. 
This led to introduce the negativity (or the logarithmic negativity $\mathcal{E}$) as follows and it has been proved that it is a good measure of bipartite entanglement for mixed states \cite{vw-02, neg-mon}.
Denoting by $|e_i^{(1)}\rangle$ and $|e_j^{(2)}\rangle$ two arbitrary bases 
in the Hilbert spaces corresponding to $A_1$ and $A_2$, the partial transpose of  $\rho_A$ with respect to $A_2$ degrees of freedom is defined as
\be 
\langle e_i^{(1)} e_j^{(2)}|\rho^{T_2}_A|e_k^{(1)} e_l^{(2)}\rangle=\langle e_i^{(1)} e_l^{(2)}|\rho_A| e^{(1)}_k e^{(2)}_j\rangle\,.
\ee
Then, the logarithmic negativity is given by 
\be
\label{neg replica limit}
{\cal E}\equiv\log ||\rho^{T_2}_A||=\log \Tr |\rho^{T_2}_A|\,,
\ee
where the trace norm  $||\rho^{T_2}_A||$ is
the sum of the absolute values of the eigenvalues of $\rho^{T_2}_A$.

A method to compute the negativity in quantum field theories based on a replica trick has been developed in \cite{cct-neg-letter, cct-neg-long}.
By observing that the integer moments $\Tr (\rho^{T_2}_A)^n$ of the partial transpose admit  different analytic continuations from even and odd integer $n$'s (here denoted by $n_e$ and $n_o$ respectively), the logarithmic negativity can be computed as the following replica limit 
\be
\label{neg replica limit}
\mathcal{E} = \lim_{n_e \to 1} \log \Tr \big(\rho_A^{T_2}\big)^{n_e} ,
\ee
taken on the sequence of moments characterised by even $n$'s. 

In the scaling limit of a critical lattice model, the moments $\Tr \rho_A^n $ of the reduced density matrix for two disjoint intervals can be computed as the four point function of peculiar fields (twist fields) or, equivalently, as the partition function of the CFT model on a particular genus $n-1$ Riemann surface $\mathcal{R}_n$ obtained through the replica procedure \cite{cct-09, cct-11, cg-08, fps-08} (in Fig.\,\ref{fig:cage} below we shown an example of $\mathcal{R}_4$).
The final result for $\Tr \rho_A^n $ when $A=A_1 \cup A_2$ encodes all the information about the underlying model, namely the central charge, the conformal dimensions of all the primaries and all the OPE coefficients \cite{cct-11, headrick}.

The expressions for $\Tr \rho_A^n $ for any $n$ are known analytically only for few CFT models: the free fermion \cite{ch-05}, the compactified (and non compactified) free boson \cite{cct-09}  and the Ising model \cite{cct-11}. 
The moments for the modular invariant Dirac fermion can be found from the ones of the compactified free boson at a specific value of the compactification radius.
These results have been confirmed by many numerical checks through the corresponding lattice models \cite{fps-08, ip-10, atc-10, fc-10, atc-11, cmv-11}. 
For the models mentioned above, $\Tr \rho_A^n $  have been written also for a generic number of disjoint intervals \cite{ctt-14}.
Since performing the replica limit $n\to 1$ of the Renyi entropies (\ref{Sndef}) for these analytic expressions is a very difficult task,  in \cite{dct-15} numerical extrapolations of the CFT analytic expressions have been done by employing the method suggested in \cite{ahjk-14}, finding excellent agreement with the corresponding lattice results. 

In this manuscript we focus on the CFT fermionic models given by the modular invariant Dirac fermion and the Ising model, which are the scaling limit of the XX spin chain and of the Ising model at the critical point, respectively.
The modular invariant partition functions on the genus $n-1$ Riemann surface $\mathcal{R}_n$ providing the moments $\Tr \rho_A^n $ for these models are written as sums over all the possible boundary conditions for the underlying fermion (either periodic or antiperiodic) around the cycles of a canonical homology basis on $\mathcal{R}_n$.
Each term in these sums is the partition function of the model on $\mathcal{R}_n$ with a fixed set of boundary conditions, i.e. with a given {\it spin structure}.
This is well known from the old days of string theory, where the bosonization on higher genus Riemann surfaces has been studied \cite{bosonization higher genus}.

As for the negativity, the replica limit approach of \cite{cct-neg-letter,cct-neg-long} has been employed to study one-dimensional conformal field theories (CFT) 
in the ground state \cite{cct-neg-letter,cct-neg-long,ctt-13, a-13, rr-14, fcd-15}, in thermal state \cite{cct-neg-T,ez-14}, in 
non-equilibrium protocols \cite{ez-14,ctc-14,hd-15,wcr-15} and for topological systems \cite{c-13,lv-13, kor}.
It is worth remarking that performing the replica limit (\ref{neg replica limit}) for the logarithmic negativity is usually more difficult than the replica limit providing the entanglement entropy from the R\'enyi entropies. 
Besides these analytical studies, the negativity has been computed numerically in several papers for various systems
\cite{wmb-09,aw-08,fcga-08,Neg3,sod, AlbaLauchlin-neg}.

In this manuscript we are interested in the entanglement between two disjoint intervals both on the lattice and in the scaling limit. 
According to \cite{cct-neg-letter,cct-neg-long}, the moments $\Tr (\rho_A^{T_2})^{n}$ in $1+1$ quantum field theories can be evaluated as four point functions of the twist fields mentioned above in a particular order or, equivalently, as the partition functions of the model on a particular genus $n-1$ Riemann surface $\widetilde{\mathcal{R}}_n$ which is different from $\mathcal{R}_n$ when $n\geqslant 3$.
Analytic expressions for  $\Tr (\rho_A^{T_2})^{n}$ are known for few simple models:  the compactified (and non compactified) free boson \cite{cct-neg-long}, the Ising model \cite{ctt-13, a-13} and the free fermion \cite{ctc-15-2}.
Again, the moments $\Tr (\rho_A^{T_2})^{n}$ for the modular invariant Dirac fermion can be found by specialising the ones of the compact boson to a particular value of the compactification radius. 
Restricting our attention to the fermionic systems given by the modular invariant Dirac fermion and the Ising model, $\Tr (\rho_A^{T_2})^{n}$ are written as sums over all the possible spin structures, similarly to the moments of the reduced density matrix.

As for the corresponding expressions on the lattice for the XX critical spin chain, the critical Ising model and the free fermion, they have been computed in \cite{ctc-15, ctc-15-2} by employing the results of \cite{ez-15} about the partial transpose of Gaussian states. 
These moments can be written as sums of computable terms. Nevertheless, this can be done for any given $n$ and a closed expression which holds at any order is not known.

In this manuscript we investigate the terms entering  in $\Tr \rho_A^n$ or $\Tr (\rho_A^{T_2})^{n}$ for the fermionic systems mentioned above, both on the lattice and in the scaling limit. 
Our main goal is to identify the scaling limit of the various terms entering in the lattice formulas for the $n$-th moment, finding that it is given by the partition function of the underlying fermionic model on $\mathcal{R}_n$ or $\widetilde{\mathcal{R}}_n$ with a given spin structure. 
This analysis allows us also to recover the corresponding results for the free fermion \cite{ch-05, ctc-15-2}.

The manuscript is organised as follows.
In \S\ref{sec:rev lattice} we introduce the spin chain models on the lattice that we are going to consider, namely the critical XX spin chain and the Ising spin chain at criticality, reviewing also  the corresponding results for the moments $\Tr \rho_A^n$ \cite{fc-10} and $\Tr (\rho_A^{T_2})^{n}$ \cite{ctc-15}.
In \S\ref{sec:cft rev} the CFT expressions for the moments of the reduced density matrix and of its partial transpose are briefly reviewed \cite{cct-neg-long, cct-11, ctt-14, ctt-13}.
In \S\ref{sec:scaling reg} we employ the fermionic coherent states formalism to identify the scaling limits of the various terms entering in the lattice expressions for $\Tr \rho_A^n$ and $\Tr (\rho_A^{T_2})^{n}$. 
In \S\ref{sec: numerics xy} we provide numerical evidences of our results for $n=2$ and $n=3$.
In \S\ref{sec:conclusions} we draw some conclusions and in \S\ref{sec:free fermion} we show that the results for $\Tr \rho_A^n$ and $\Tr (\rho_A^{T_2})^{n}$ for the free fermion, found in \cite{ch-05} and \cite{ctc-15-2} respectively, can be recovered from the corresponding expressions for the critical XX spin chain on the lattice or for the modular invariant Dirac fermion in the scaling limit. 
\\

\section{Review of the lattice results}
\label{sec:rev lattice}

In this manuscript we consider the lattice models given by the XX spin chain at criticality  and by the critical Ising spin chain.
Given a subsystem $A$ made by two disjoint blocks,  in this section  we review the lattice computations of the moments of the reduced density matrix $\Tr \rho_A^n $ \cite{fc-10} and of its partial transpose $\Tr (\rho^{T_2}_A)^n $ \cite{ctc-15}.

\subsection{Hamiltonians}
\label{sec: lattice H}

The Hamiltonians of the XX spin chain at the critical point and of the critical Ising spin chain read respectively 
\be
\label{eq:HXX Hising}
H_\text{XX} = -\frac{1}{4} 
\sum_{j=1}^L \left(
\sigma_j^x \sigma_{j+1}^x 
+ \sigma_j^y \sigma_{j+1}^y
\right)\,,
\qquad
H_{\text{Ising}} = -\frac{1}{2} 
\sum_{j=1}^L \left(
 \sigma_j^x \sigma_{j+1}^x 
+ \,\sigma_j^z
\right)\,,
\ee
where $\sigma_{j}^\alpha$ (with $\alpha \in \{x,y,z\}$) are the Pauli matrices at the $j$-th site of the chain with $L$ sites and periodic boundary conditions $\sigma_{L+1}^\alpha=\sigma_1^\alpha $ have been assumed. 
It is well known that the Hamiltonians (\ref{eq:HXX Hising}) can be diagonalized by employing the Jordan-Wigner transformation
\be
\label{JW map}
c_j = 
\Big( \prod_{m<j} \sigma_m^z \Big) 
\frac{\sigma_j^x - \textrm{i} \sigma_j^z}{2}\,,
\qquad
c_j^\dagger = 
\Big( \prod_{m<j} \sigma_m^z \Big) 
\frac{\sigma_j^x + \textrm{i} \sigma_j^z}{2}\,,
\ee
which maps the spin variables into anti-commuting fermionic variables (i.e.\ $\{c_i, c^\dagger_j\}=\delta_{ij}$).
In terms of the latter fermionic variables, the Hamiltonians in \eqref{eq:HXX Hising} become respectively
\begin{subequations}
\bea
\label{eq:HXX}
& &
H_\text{XX} = \frac12 \sum_{i=1}^L \left( 
c^\dagger_i c_{i+1}+ c^\dagger_{i+1} c_{i}\right),
\\
\label{eq:Hising}
& &
H_\text{Ising}=\sum_{i=1}^L \left(\frac12 \left[c^\dagger_i c^\dagger_{i+1}+ c_{i+1} c_{i}
+c^\dagger_i c_{i+1}+ c^\dagger_{i+1} c_{i} \right]-  c^\dagger_i c_i
\right) ,
\eea
\end{subequations}
where boundary and additive terms have been discarded.
Since the Hamiltonians in \eqref{eq:HXX} and \eqref{eq:Hising}  are quadratic, they can be easily diagonalised in the momentum space through a Bogoliubov transformation.

In order to study the reduced density matrices associated to spin blocks, it is very useful to introduce also the following Majorana fermions \cite{vlrk-03}
\begin{equation}
\label{a from c def}
a_{j}^x = c_j + c_j^\dagger \,,
\qquad
a_{j}^y = \iu (c_j - c_j^\dagger) \,,
\end{equation}
which satisfy the anticommutation relations $\{a^\alpha_r, a^\beta_s\}=2\delta_{\alpha\beta}\delta_{rs}$.
Moreover, given a block $C$ of contiguous sites, a crucial operator we need in our analysis is
\begin{equation}
\label{P_C def}
 P_{C} = \prod_{j\in C} \iu \, a^x_j a^y_j \,.
\end{equation}
This string of Majorana operators satisfies $P_C^{-1} =P_C$.

\subsection{Moments of the reduced density matrix}
\label{subsec rho_A}

In a spin $1/2$ chain, the  reduced density matrix $\rho_A =\Tr_B |\Psi \rangle \langle \Psi |$ of 
$A=A_1\cup A_2$ can be computed by summing all the operators in $A$ as follows \cite{vlrk-03}
\be
\label{rhoA correlators}
\rho_A = \frac{1}{2^{\ell_1+\ell_2}} \sum_{\nu_j} 
\Big\langle \prod_{j \in A} \sigma_j^{\nu_j} \Big\rangle 
 \prod_{j \in A} \sigma_j^{\nu_j} \,,
\ee
where $\nu_j \in \{ 0,1,2,3 \}$, being $\sigma^{0} = {\bf 1}$  
the identity matrix and $\sigma^{1} = \sigma^{x}$, $\sigma^{2} = \sigma^{y}$, $\sigma^{3} = \sigma^{z}$ the Pauli matrices. 
For a generic chain, the correlators in (\ref{rhoA correlators}) are very difficult to evaluate.
Nevertheless, when the state can be written in terms of free fermions, the Wick theorem can be employed to find them.

For the single interval case, the Jordan-Wigner transformation  (\ref{JW map})  sends the first 
$\ell$ spins into the first $\ell$ fermions, hence the spin and fermionic entropy coincide \cite{vlrk-03}.
Unfortunately, this is not the case when $A$ is made by two disjoint blocks because also 
the fermions in the block $B_1$ separating them (see Fig.\,\ref{fig:sc}) contribute to the spin reduced density matrix of $A_1\cup A_2$ \cite{atc-10,ip-10,fc-10}. 

Focussing on the models we are interested in, since the Hamiltonians (\ref{eq:HXX Hising}) commute with $\prod_{j=1}^L \sigma^z_j$, 
the expectation values of operators containing an odd number of fermions vanish in (\ref{rhoA correlators}).
Thus, since the total number of fermions in $A_1\cup A_2$ must be even, the numbers of fermions in $A_1$ and $A_2$ are either both even or both odd.
This means that the spin reduced density matrix $\rho_A$ of $A = A_1 \cup A_2$ can be written as follows \cite{fc-10}
\be
\label{rhoA dec even-odd}
\rho_A 
\,=\, 
\rho_{\rm even} +  P_{B_1} \, \rho_{\rm odd} \,,
\ee
where $P_{B_1}$ is the string of Majorana operators (\ref{P_C def}) associated to the block $B_1$ and
\begin{subequations}
\bea
\label{rho even def}
& &
\rho_{\rm even} \equiv
\frac{1}{2^{\ell_1+\ell_2}}
\sum_{\rm even}
w_{12}\, O_1 O_2 \,,
\hspace{1.5cm}
w_{12} \equiv  \langle O_2^\dagger O_1^\dagger  \rangle \,,
\\
\label{rho odd def}
& &
\rho_{\rm odd}  \equiv
\frac{\langle P_{B_1} \rangle}{2^{\ell_1+\ell_2}}
\sum_{\rm odd}
w_{12}^{B_1}\, O_1 O_2 \,,
\hspace{1.5cm}
w^{B_1}_{12} \equiv  \frac{\langle O_2^\dagger P_{B_1} O_1^\dagger  \rangle}{\langle P_{B_1} \rangle} \,,
\eea
\end{subequations}
being $O_k$ (with $k\in \{1,2\}$) an arbitrary product of Majorana fermions in $A_k$, namely $O_k = \prod_{j\in A_k}(a^x_{j})^{\mu^x_{[j]}}(a^y_{j})^{\mu^y_{[j]}}$ with $\mu_{[j]}^\alpha\in \{0,1\}$.
The notation $\sum_{\rm even}$ ($\sum_{\rm odd}$) indicates that the sum is restricted to operators $O_k$ containing an even (odd) number of Majorana fermions.

In order to compute the moments $\Tr\rho_A^n$ for the ground state $|\Psi \rangle \langle \Psi |$ of the spin models defined in (\ref{eq:HXX Hising}), it is convenient to introduce both the fermionic reduced density matrix of $A=A_1 \cup A_2$, namely\footnote{The superscript $\boldsymbol 1$ distinguishes the fermionic density matrix from the spin density matrix $\rho_A$.}
\be
\label{rhoA ff}
\rho_A^{\,{\bf 1}}
\equiv
\frac{1}{2^{\ell_1+\ell_2}}
%\sum_{\substack{ \rm even} \\  {\rm odd}}
\sum_{{\rm even} \atop {\rm odd}}
w_{12}\, O_1 O_2 \,,
\ee
(we recall that $\langle O_1 O_2 \rangle$ vanishes when the numbers of fermionic operators in $O_1$ and $O_2$ have different parity) and the following auxiliary density matrix
\be
\label{fake rhoA}
\rho_A^{B_1} 
\equiv \frac{\Tr_B \big(P_{B_1}  |\Psi \rangle \langle \Psi | \big)}{\langle P_{B_1}  \rangle}
=
\frac{1}{2^{\ell_1+\ell_2}}
%\sum_{\substack{ \rm even} \\  {\rm odd}}
\sum_{{\rm even} \atop {\rm odd}}
w_{12}^{B_1}\, O_1 O_2 \,,
\ee
which satisfies the normalisation condition $\Tr  \rho_A^{B_1} =1 $. 
It is worth remarking that $\rho_A = \rho_A^{\,{\bf 1}} $ for the free fermion \cite{pascazio-08}, as one can easily observe  by setting $P_{B_1} = {\boldsymbol 1}$ in \eqref{rhoA dec even-odd}.

By using that $P_{A_2} a_j^\alpha P_{A_2}$ gives either $- a_j^\alpha $ or $a_j^\alpha $, depending on whether $j\in A_2$ or $j\notin A_2$ respectively, for (\ref{rhoA ff}) one finds that
\be
\label{PA2 rho ff}
P_{A_2} \rho_A^{\,{\bf 1}} P_{A_2}
=
\frac{1}{2^{\ell_1+\ell_2}}
%\sum_{\substack{ \rm even} \\  {\rm odd}}
\sum_{{\rm even} \atop {\rm odd}} (-1)^{\mu_2}
w_{12}\, O_1 O_2 \,,
\ee
where $\mu_2 = \sum_{j \in A_2} (\mu_{[j]}^x + \mu_{[j]}^y)$ is the total number of Majorana operators occurring in $O_2$.
An expression similar to (\ref{PA2 rho ff}) can be written also for (\ref{fake rhoA}) and, from these results, it is straightforward to conclude that the operators in (\ref{rho even def}) and (\ref{rho odd def}) can be written as \cite{fc-10} 
\be
\label{rho pm ferm}
\rho_{\rm even}   =
\frac{\rho_A^{\,{\bf 1}} + P_{A_2}  \rho_A^{\,{\bf 1}} P_{A_2} }{2}\,,
\qquad
\rho_{\rm odd}  =
\langle P_{B_1} \rangle
\,\frac{\rho_A^{B_1} - P_{A_2}  \rho_A^{B_1} P_{A_2} }{2}\,.
\ee
Considering the four fermionic Gaussian operators  occurring in the r.h.s.'s of (\ref{rho pm ferm}), let us introduce the following notation 
\begin{equation}
\label{rho 1234 def}
\rho_1 \equiv \rho_A^{\,{\bf 1}} \,,
\qquad
\rho_2 \equiv P_{A_2}  \rho_A^{\,{\bf 1}} P_{A_2} \,,
\qquad
\rho_3 \equiv  \langle P_{B_1} \rangle\, \rho_A^{B_1} \,,
\qquad
\rho_4 \equiv  \langle P_{B_1} \rangle\, P_{A_2}  \rho_A^{B_1} P_{A_2} \,.
\end{equation}
In terms of these four matrices, from (\ref{PA2 rho ff}) one finds that the reduced density matrix (\ref{rhoA dec even-odd})  can be written as
\be
\label{rhoA 1234}
\rho_A 
\,=\, \frac{1}{2} \Big[ \,
\rho_1 + \rho_2 + P_{B_1} \big( \rho_3 - \rho_4  \big)
\Big] \,.
\ee

In \cite{fc-10} it has been shown that 
\be
\label{renyi rho plus}
\Tr \rho_A^n = \Tr (\rho_{\rm even} + \rho_{\rm odd})^n
=
\frac{1}{2^n} \, \Tr \big( \rho_1 + \rho_2 + \rho_3 - \rho_4 \big)^n \,,
\ee
where in the last step (\ref{rho pm ferm}) and (\ref{rho 1234 def}) have been employed.
Eq.\,(\ref{renyi rho plus}) tells us that $\textrm{Tr} \rho_A^n $ is a sum of $4^n$ terms, where each term is characterised by a string $\boldsymbol{q}$ made by $n$ elements $q_i \in \{1,2,3,4\}$.
Once the explicit expressions (\ref{rho 1234 def}) are plugged into such sum, the terms with an odd number of 4's in $\boldsymbol{q}$ occur in $\Tr \rho_A^n $ with a minus sign.
By  using the cyclic  property of the trace and the fact that $P_{A_2}^2={\bf 1}$, it is straightforward to observe that a term characterised by $\boldsymbol{q}$ is equal to the one characterised by $\boldsymbol{q}'$, with  $\boldsymbol{q}'$ obtained by exchanging $1\leftrightarrow 2$ and $3\leftrightarrow 4$ in $\boldsymbol{q}$.
Moreover, a term associated to a $\boldsymbol q$ with an odd total number of 3's and 4's will have opposite sign with respect to the corresponding one characterised by $\boldsymbol q'$, and they cancel out. Instead, a term having a $\boldsymbol q$ with an even total number of 3's and 4's has the same sign of the corresponding one given by $\boldsymbol q'$, and this provide a factor $2$ that can be collected.

After these simplifications, the net result  is
\begin{equation}
\label{renyi T_n}
\textrm{Tr}\rho_A^n
=
\frac{1}{2^{n-1}}\,
\sum_{\boldsymbol{q}} 
(-1)^{\#4}\,
 \Tr\bigg[ \, \prod_{k=1}^{n} \rho_{q_k} \bigg] ,
\end{equation}
where the sum is over all $\boldsymbol q$'s with an even total number of 3's and 4's and modulo the exchange $1\leftrightarrow 2$ and $3\leftrightarrow 4$.
It is not difficult to realize that  the sum \eqref{renyi T_n}  contains $2^{2(n-1)}$ terms and the ones having an odd number of $\rho_4$'s occur with a minus sign.

Each term in the sum (\ref{renyi T_n}) enjoys a $\mathbb{Z}_n \times \mathbb{Z}_2$ symmetry (dihedral symmetry).
The $\mathbb{Z}_n$ symmetry comes from the fact that we can permute cyclically the $n$ factors within the trace.
The $\mathbb{Z}_2$ invariance occurs because each term in (\ref{renyi T_n})  is real and every factor $\rho_{q_k}$ is hermitian.
These observations imply that every term in (\ref{renyi T_n})  can be written by taking all the factors within the trace in the opposite order. 
Taking into account the dihedral symmetry, many terms in the sum (\ref{renyi T_n}) coincide and therefore the moments $\textrm{Tr}\rho_A^n$ can be written as linear combinations with integer coefficients of a minimal number of representative terms belonging to different equivalence classes identified by the dihedral symmetry.

Explicit expressions of \eqref{renyi T_n} for $2\leqslant n \leqslant 5$ have been written in \cite{fc-10, ctc-15} in terms of the correlators. 
By employing the matrices (\ref{rho 1234 def}), the simplest cases of $n=2$ and $n=3$ are given by
\bea
\label{eq:TrrhoA n2}
& & \hspace{-1.5cm}
\textrm{Tr}\rho_A^2
\;=\;
\frac{1}{2}\Big[\,
\Tr(\rho_1^2)+\Tr(\rho_1 \rho_2)+\Tr(\rho_3^2)-\Tr(\rho_3\rho_4)
\,\Big] \,,
\\
\label{eq:TrrhoA n3}
\rule{0pt}{.8cm}
& & \hspace{-1.5cm}
\textrm{Tr}\rho_A^3
\;=\;
\frac{1}{4}\Big[\,
\Tr(\rho_1^3)
+3\,\Tr(\rho_1^2 \rho_2) +3\,\Tr(\rho_1\rho_3^2)+3\,\Tr(\rho_2\rho_3^2)
-6\,\Tr(\rho_1\rho_4\rho_3 )\,\Big]  \,.
\eea

In \cite{ctc-15} the partial transpose $\rho_A^{T_2}$ for (\ref{rhoA dec even-odd}) has been studied by employing the previous analysis for $\rho_A$ and the prescription for the partial transpose of a fermionic Gaussian operator introduced in \cite{ez-15}.

\subsection{Moments of the partial transpoe}
\label{subsec pt2 lattice}

Given a Gaussian density matrix written in terms of Majorana fermions in $A=A_1 \cup A_2$, the partial transposition 
with respect to $A_2$ acts only on the modes in $A_2$, leaving invariant the ones in $A_1$.
Furthermore, also the operator $P_{B_1}$ is left unchanged.
Since the partial transposition is linear, from \eqref{rhoA dec even-odd} we have
\begin{equation}
\label{rhoA T2}
\rho_A^{T_2} 
\,=\, 
\rho_{\rm even}^{T_2} +  P_{B_1} \, \rho_{\rm odd}^{T_2} \,.
\end{equation}

Considering the operator $O_2$ made by Majorana fermions in $A_2$, its partial transposition is given by \cite{ez-15} 
\begin{equation}
\label{hat transposition def}
 O_2^{T} \,=\, (-1)^{\tau(\mu_2)} O_2 \,,
\end{equation}
where
\begin{equation}
\label{tau mu2}
 \tau(\mu_2) = 
 \begin{cases}
  0 & (\mu_2 \textrm{ mod } 4) \in \{0,1\}  \,, \\
  1 & (\mu_2 \textrm{ mod } 4) \in \{2,3\}  \,.
 \end{cases}
\end{equation}
The partial transposition can be defined also in another way, which is related to (\ref{hat transposition def}) through a unitary transformation \cite{ctc-15-2}.

By applying \eqref{hat transposition def} to (\ref{rho even def}) and (\ref{rho odd def}), we find respectively
\be
\label{rho T2 even-odd}
\rho_{\rm even}^{T_2} =
\frac{1}{2^{\ell_1+\ell_2}}
\sum_{\rm even}
(-1)^{\mu_2/2}\,
w_{12} \, O_1 O_2 \,,
\qquad
\rho_{\rm odd}^{T_2} =
\frac{\braket{P_{B_1}}}{2^{\ell_1+\ell_2}}
\sum_{\rm odd}
(-1)^{(\mu_2-1)/2}\,
w_{12}^{B_1}\, O_1 O_2 \,.
\ee
Plugging these expressions into \eqref{rhoA T2}, one gets the partial transpose of the spin reduced density matrix in terms of the Majorana fermions. 
Both $\rho_\text{even}^{T_2}$ and $\rho_\text{odd}^{T_2}$ in (\ref{rho T2 even-odd}) can be written as a sum of two Gaussian matrices.
Indeed, by introducing
\begin{equation}
\label{rhoA tilde B1 def}
\tilde{\rho}_A^{\,{\bf 1}} \equiv
\frac{1}{2^{\ell_1+\ell_2}}
%\sum_{\substack{{\rm even} \\ {\rm odd}}}
%\sum_{{\rm even} \atop {\rm odd}}
\sum_{{\rm even} \atop {\rm odd}}
\iu^{\mu_2}\,
w_{12}
\, O_1 O_2 \,,
\qquad
\tilde{\rho}_A^{B_1} \equiv
\frac{1}{2^{\ell_1+\ell_2}}
%\sum_{{\rm even} \atop {\rm odd}}
%\sum_{\substack{{\rm even} \\ {\rm odd}}}
\sum_{{\rm even} \atop {\rm odd}}
\iu^{\mu_2}\,
w_{12}^{B_1}
\, O_1 O_2 \,,
\end{equation}
the matrices in (\ref{rho T2 even-odd}) become respectively
\begin{equation}
\label{rhoT2 even-odd}
\rho_{\rm even}^{T_2} =
\frac{\tilde{\rho}_A^{\,{\bf 1}} + P_{A_2}  \tilde{\rho}_A^{\,{\bf 1}} P_{A_2} }{2} \,,
\qquad
\rho_{\rm odd}^{T_2}  =
\langle P_{B_1} \rangle\,
\frac{\tilde{\rho}_A^{B_1} - P_{A_2}  \tilde{\rho}_A^{B_1} P_{A_2} }{2 \iu} \,.
\end{equation}
While $\rho_{\rm even}^{T_2}$ and $\rho_{\rm odd}^{T_2} $ in (\ref{rhoT2 even-odd}) are Hermitian, the matrices in (\ref{rhoA tilde B1 def}), which are used to build them, are not.
Indeed, $(\tilde{\rho}_A^{\,{\bf 1}})^\dagger  = P_{A_2}  \tilde{\rho}_A^{\,{\bf 1}} P_{A_2}$ 
and $(\tilde{\rho}_A^{B_1} )^\dagger = P_{A_2}  \tilde{\rho}_A^{B_1} P_{A_2}$.

Mimicking the analysis performed above for the reduced density matrix, we find it convenient to introduce the following four fermionic Gaussian operators
\begin{equation}
\label{rho tilde 1234 def}
\tilde{\rho}_1 \equiv \tilde{\rho}_A^{\,{\bf 1}} \,,
\qquad
\tilde{\rho}_2 \equiv P_{A_2}  \tilde{\rho}_A^{\,{\bf 1}} P_{A_2} \,,
\qquad
\tilde{\rho}_3 \equiv  \langle P_{B_1} \rangle\, \tilde{\rho}_A^{B_1} \,,
\qquad
\tilde{\rho}_4 \equiv  \langle P_{B_1} \rangle\, P_{A_2}  \tilde{\rho}_A^{B_1} P_{A_2} \,.
\end{equation}
In terms of these matrices,  from (\ref{rhoA T2}) and (\ref{rhoT2 even-odd}) we have that the partial transpose of the reduced density matrix can be written as
\be
\label{rhoA T2 1234}
\rho_A^{T_2} 
\,=\, \frac{1}{2} \Big[ \,
\tilde{\rho}_1 + \tilde{\rho}_2  -\iu\, P_{B_1} \big( \tilde{\rho}_3 - \tilde{\rho}_4  \big)
\Big] \,.
\ee

The moments of the partial transpose can be computed as follows \cite{ctc-15}
\begin{equation}
\label{traces n rhopm T2}
\Tr \big( \rho_A^{T_2}\big)^n 
=  \Tr \big(\rho_\text{even}^{T_2}+\rho_\text{odd}^{T_2}\big)^n
= \frac{1}{2^n} \, \Tr \big( \tilde{\rho}_1 + \tilde{\rho}_2 -\iu\, \tilde{\rho}_3 + \iu\,\tilde{\rho}_4 \big)^n\,,
\end{equation}
where in the last step (\ref{rhoT2 even-odd}) and (\ref{rho tilde 1234 def}) have been used. 
Thus,  $\Tr\big(\rho_A^{T_2}\big)^n$ in  \eqref{traces n rhopm T2} turns out to be a linear combination of $4^n$ terms, where each term is specified by a string $\tilde{\boldsymbol{q}}$ made by $n$ elements $\tilde{q}_i \in \{1,2,3,4\}$.
This result is similar to (\ref{renyi rho plus}) obtained for the moments of the reduced density matrix.
An important difference between \eqref{rho 1234 def} and \eqref{rho tilde 1234 def} is the occurrence of the imaginary unit in (\ref{traces n rhopm T2}), which implies that the sign in front of each term is $(-1)^{\#3}\,\iu^{\#3+\#4}$, being $\#3$ and $\#4$ the number of $3$'s and $4$'s respectively occurring in $\tilde{\boldsymbol{q}}$.

The analysis of the various terms occurring in the sum given by (\ref{traces n rhopm T2}) is very similar to the one performed for $\Tr \rho_A^n$.
In particular, a term characterised by a vector $\tilde{\boldsymbol{q}}$ is equal to the one associated to the vector $\tilde{\boldsymbol{q}}'$ obtained by exchanging $1\leftrightarrow 2$ and $3\leftrightarrow 4$ and therefore the terms whose $\tilde{\boldsymbol{q}}$ has an odd total number of $\tilde{\rho}_3$'s and $\tilde{\rho}_4$'s cancel out because of the opposite relative sign between  $\tilde{\rho}_3$ and $\tilde{\rho}_4$ in (\ref{traces n rhopm T2}), while the ones having an even total number of $\tilde{\rho}_3$'s and $\tilde{\rho}_4$'s sum pairwise and a factor of $2$ can be collected out. 
Since $\#3+\#4$ is even, all the non vanishing coefficients in $\Tr\big(\rho_A^{T_2} \big)^n $ are real and their overall sign is $(-1)^{\frac{\#3+\#4}{2} + \#3}$.

These observations allow us to write the moments of the partial transpose $\rho_A^{T_2} $ as 
\begin{equation}
\label{ptA2 tilde T_n}
\Tr \big( \rho_A^{T_2} \big)^n 
=
\frac{1}{2^{n-1}}\,
\sum_{\tilde{\boldsymbol{q}}}
%\, \tilde{v}_n(\tilde{\boldsymbol{q}}) \,
(-1)^{\frac{\#4-\#3}{2}}\,
 \Tr\bigg[ \, \prod_{k=1}^{n} \tilde{\rho}_{\tilde q_k} \bigg] ,
\end{equation}
where, like in (\ref{renyi T_n}), the sum is assumed to be over all $\tilde{\boldsymbol q}$ with an even total number of 3's and 4's and modulo the exchange $1\leftrightarrow 2$ and $3\leftrightarrow 4$.
Thus, the sum (\ref{ptA2 tilde T_n}) contains $2^{2(n-1)}$ terms.

Each term in the sum (\ref{ptA2 tilde T_n}) enjoys the dihedral symmetry $\mathbb{Z}_n \times \mathbb{Z}_2$.
The $\mathbb{Z}_n$ invariance comes from the cyclic permutation of the factors within each trace, like for the terms in (\ref{renyi T_n}),
while the $\mathbb{Z}_2$ symmetry is related to the reality of each term.
As already pointed out below (\ref{rhoT2 even-odd}), the four terms in \eqref{rho tilde 1234 def} are not separately hermitian.
The hermitian conjugation exchanges $\tilde\rho_1 \leftrightarrow\tilde\rho_2$ and $\tilde\rho_3 \leftrightarrow\tilde\rho_4$, and any term in the sum \eqref{ptA2 tilde T_n} is left invariant under such exchange, as already remarked above.
By employing the above behaviour of the matrices $\tilde\rho_j$ under hermitian conjugation and the fact that $P_{B_1}$ is hermitian, one concludes that also the matrix $\rho_A^{T_2}$ in (\ref{rhoA T2 1234}) is hermitian.

We find it useful to report explicitly \eqref{ptA2 tilde T_n} at least in the simplest cases of $n=2$ and $n=3$.
They read respectively \cite{ctc-15}
\bea
\label{eq:TrrhoAT n2}
& & \hspace{-1.5cm}
\Tr \big( \rho_A^{T_2} \big)^2
\;=\;
\frac{1}{2}\Big[\,
\Tr (\tilde\rho_1^2)+\Tr (\tilde\rho_1\tilde\rho_2)+\Tr(\tilde\rho_3\tilde\rho_4)-\Tr(\tilde\rho_3^2)
\,\Big] \,,
\\
\label{eq:TrrhoAT n3}
\rule{0pt}{.8cm}
& & \hspace{-1.5cm}
\Tr \big( \rho_A^{T_2} \big)^3
\;=\;
\frac{1}{4}\Big[\,
\Tr(\tilde\rho_1^3)
+3\,\Tr(\tilde\rho_1^2 \tilde\rho_2)
+6\,\Tr(\tilde\rho_1\tilde\rho_4\tilde\rho_3)
-3\,\Tr(\tilde\rho_1\tilde\rho_3^2)
-3\,\Tr (\tilde\rho_2\tilde\rho_3^2)
\,\Big] \,.
\eea
The formulas written in \cite{ctc-15} for $n=4$ and $n=5$ can be also expressed in terms of the matrices (\ref{rho tilde 1234 def}), but the number of terms to deal with significantly increases with respect to (\ref{eq:TrrhoAT n2}) and (\ref{eq:TrrhoAT n3}).

\section{Review of the CFT results}
\label{sec:cft rev}

The scaling limit of the lattice models considered in the previous section are described by two dimensional conformal field theories. 
In particular, the scaling limit of the critical XX spin chain is the modular invariant Dirac fermion, whose central charge is $c=1$, while for the critical Ising chain is the Ising model ($c=1/2$).
In this section we review the CFT results for the R\'enyi entropies and the moments of the partial transpose for the modular invariant Dirac fermion and the Ising model.

When the subsystem $A$ is an interval of length $\ell $ on the infinite line, the moments of the reduced density matrix for a CFT with central charge $c$
can be written as \cite{Holzhey,cc-04,cc-rev}
\be
\label{Renyi:asymp}
\Tr\rho_A^{n}= c_n \left(\frac{\ell}{a}\right)^{-c(n-1/n)/6},
\ee
where $a$ is the inverse of an ultraviolet cutoff (e.g.\ the lattice spacing) and $c_n$ is a non universal constant.
For the Dirac fermion $c=1$ and \cite{jk-04} 
\begin{equation}
\label{c_n free fermion}
 c_n^{\textrm{XX}} = 2^{-\frac{1}{6}\left(n-\frac{1}{n}\right)}
 \exp\bigg\{ \,\iu\, n\int_{-\infty}^\infty
 \log\bigg(\frac{\Gamma\left(\frac{1}{2}+\iu z\right)}{\Gamma\left(\frac{1}{2}-\iu z\right)}\bigg) 
 \big[\tanh\left(\pi z\right) - \tanh\left(\pi n z\right) \big] \mathrm{d}z\, \bigg\} \,,
\end{equation}
while for Ising model $c=1/2$ and  \cite{ij-08,ccd-09} 
\begin{equation}
\label{c_n ising}
 c_n^\text{Ising} = 2^{-\frac{1}{12}\left(n-\frac{1}{n}\right)} \sqrt{c_n^{\textrm{XX}} } \,.
\end{equation}
The expression (\ref{Renyi:asymp}) can be found by realising that  $\Tr\rho_A^{n}$ is the 
partition function on a sphere obtained by attaching cyclically the $n$ replicas \cite{cc-04}.
It  can also be interpreted as the two-point function of some twist operators acting at the endpoints $u$ and $v$ of the interval $A$
 \cite{cc-04,ccd-09}, namely $\Tr\rho_A^{n}= \langle {\cal T}_n(u)\bar {\cal T}_n(v)\rangle$. 
The twist fields ${\cal T}_n$ and $\bar {\cal T}_n$ behave like spinless primary operators whose scaling dimensions read
\be
\label{Delta_n}
\Delta_n= \frac{c}{12}\left(n-\frac1n\right) .
\ee
From  the moments $\Tr \rho_A^n$ one can find information about the full spectrum of the reduced density matrix \cite{cl-08}.

\begin{figure}
\vspace{.6cm}
\begin{center}
\includegraphics[width=.65\textwidth]{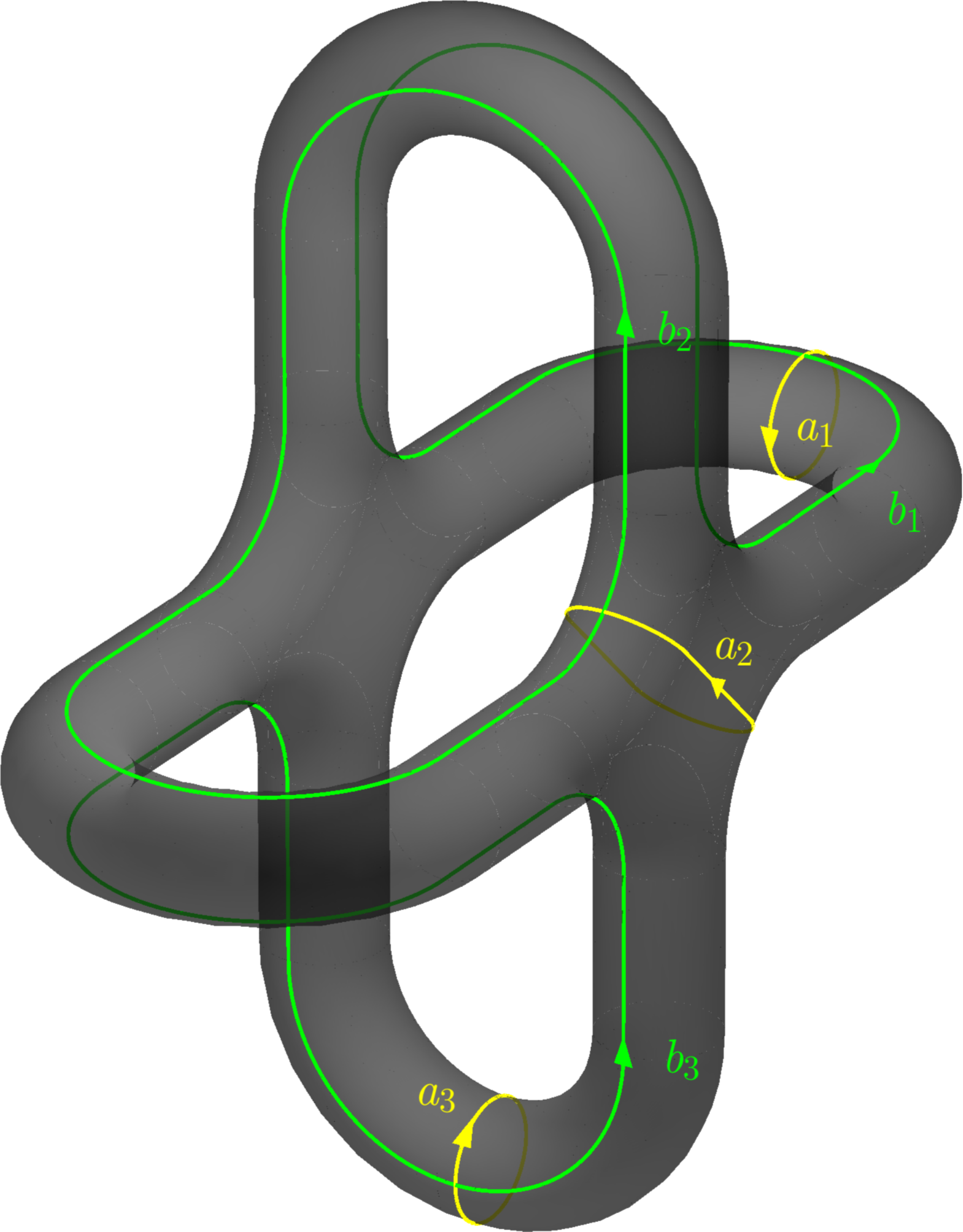}
\end{center}
\vspace{.1cm}
\caption{The Riemann surface $\mathcal{R}_4$ and the canonical homology basis considered in this manuscript (see also \cite{cct-11, ctt-14}).}
\label{fig:cage}
\end{figure}

\begin{figure}
\vspace{-.0cm}
\begin{center}
\includegraphics[width=.8\textwidth]{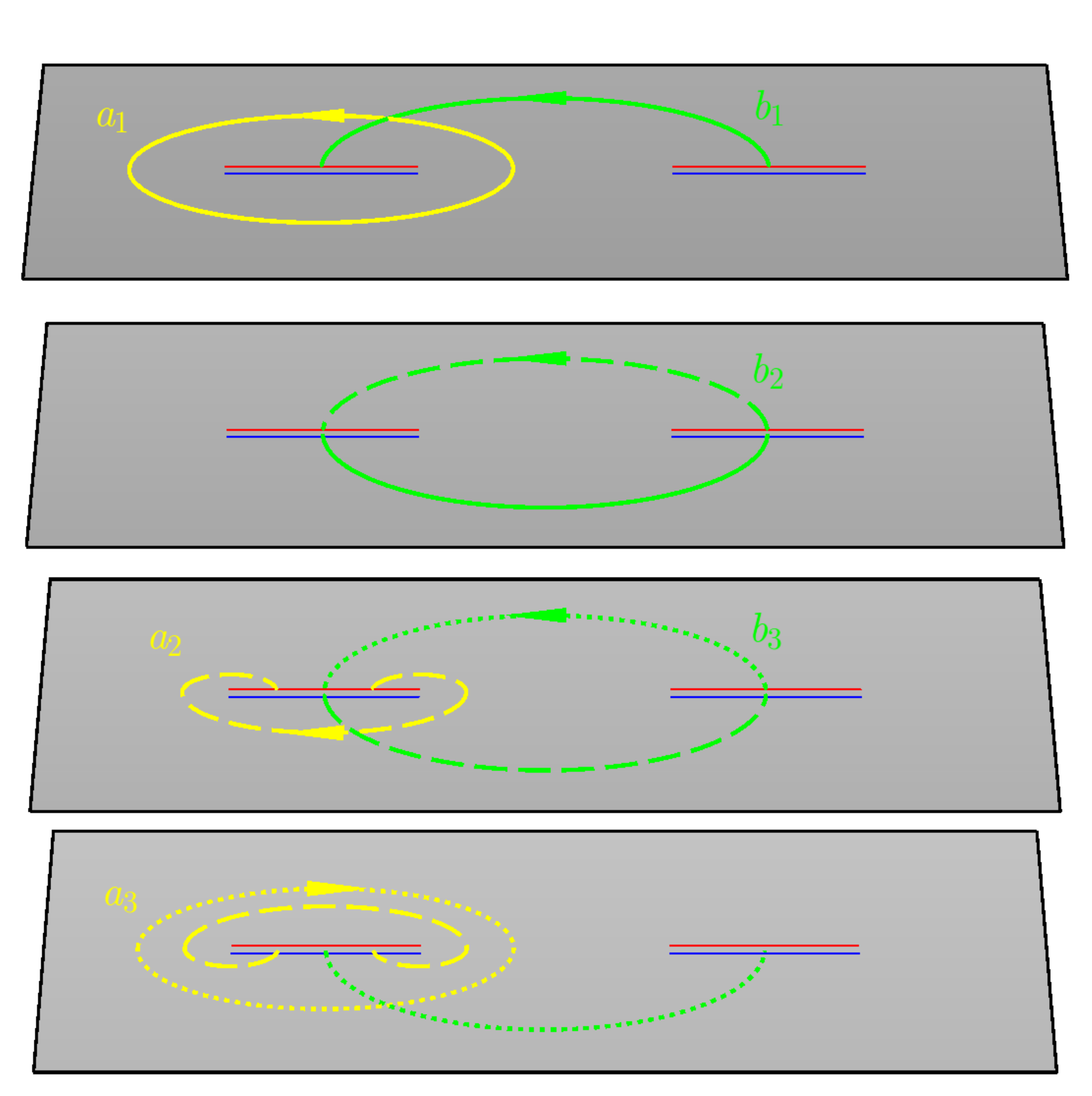}
\end{center}
\vspace{.0cm}
\caption{The Riemann surface $\mathcal{R}_4$ represented through the cyclic joining of four sheets. 
For a given slit, the upper edge (red) should be identified with the lower edge (blue) of the slit below, in a cyclic way. 
The canonical homology basis is the same one shown in Fig.\,\ref{fig:cage}, with the same colour code. }
\label{fig:R4}
\end{figure}

When $A$ is made by two disjoint intervals $A=A_1\cup A_2=[u_1,v_1]\cup [u_2,v_2]$ on the infinite line whose endpoints are ordered as $u_1<v_1<u_2<v_2$,
the moments $\Tr \rho_A^n$  are given by the four point function $\langle {\cal T}_n(u_1)\bar {\cal T}_n(v_1) {\cal T}_n(u_2)\bar {\cal T}_n(v_2) \rangle$ \cite{cc-04, cg-08, fps-08, cct-09, cct-11}.
Global conformal invariance allows to write $\Tr \rho_A^n$ as follows
(hereafter we will drop the dependence on the UV cutoff $a$)
\be
\label{Fn}
\Tr \rho_A^n
=c_n^2 \left(\frac{(u_2-u_1)(v_2-v_1)}{(v_1-u_1)(v_2-u_2)  (v_2-u_1) (v_1-u_2)} 
\right)^{2\Delta_n} \mathcal{F}_n(x) \,,
\ee 
where $c_n$ is the same non-universal constant introduced in (\ref{Renyi:asymp}), $x$ is the four point ratio 
\be
\label{cross ratio def}
x =\frac{(u_1 -v_1)(u_2-v_2)}{(u_1 - u_2)(v_1 - v_2)} \,\in\, (0,1) \,,
\ee
and the normalization $\mathcal{F}_n(0)=1$ has been imposed.
The universal function $\mathcal{F}_n(x)$ encodes all the information about 
the operator content of the CFT and it has been largely studied during the last years \cite{cct-09,cct-11,cg-08,ch-05,fps-08,c-10,headrick,hlr-13,ctt-14,atc-10,atc-11,rg-12,f-12} 
(see also \cite{RT, headrick, hol} for the holographic viewpoint and \cite{cft-high-dims} for higher dimensional conformal field theories).
The expression (\ref{Fn}) for the  moments $\Tr \rho_A^n$ of two disjoint intervals can be interpreted also as the partition function of the underlying model on the Riemann surface ${\cal R}_n$
of genus $n-1$. The Riemann surface ${\cal R}_n$ can be constructed by attaching cyclically the $n$ sheets on which the $n$ copies of the model are defined (we remark that each sheet has the topology of a cylinder) \cite{cct-11}. 
Since ${\cal R}_n$ has been obtained through a replica method, it does not represent the most generic genus $n-1$ Riemann surface.
In Fig.\,\ref{fig:cage} and Fig.\,\ref{fig:R4} we show two representations of ${\cal R}_4$, whose genus is equal to three.
In Fig.\,\ref{fig:R4} the sheets should be thought as attached in a cyclic way along the edges of the slits: the lower edge (blue) of each slit should be identified with the upper edge (red) of the slit just above.

The genus $n-1$ Riemann surface ${\cal R}_n$ can be defined as the complex curve in $\mathbb{C}^2$ given by $y^n = (z-u_1)(z-u_2) [(z-v_1)(z-v_2)]^{n-1}$ \cite{cct-09, eg-03}, being $(y,z) \in \mathbb{C}^2$.
A crucial object for our analysis is the period matrix, which is a $g \times g$ complex symmetric matrix with positive definite imaginary part for a generic genus $g$ Riemann surface \cite{bosonization higher genus}.  
In order to write the period matrix $\tau$ of the Riemann surface ${\cal R}_n$, one needs to introduce 
a canonical homology basis, namely a set $\{a_r, b_r \,; \, 1\leqslant r\leqslant n-1\}$ of closed oriented curves on ${\cal R}_n$ such that $a_i \cap a_j = b_i \cap b_j =\emptyset$ and, for any fixed $i$, the cycle $a_i $ crosses only the cycle $b_i$ in such way that the cross product of the tangent vectors at the intersection points either inward or outward.
In this manuscript we choose the canonical homology basis shown in Figs.\,\ref{fig:cage} and \ref{fig:R4}, which has been discussed in \cite{ctt-14}.

For fermionic models, it is also crucial to specify the boundary conditions (either periodic or antiperiodic) along all the cycles of a canonical homology basis \cite{bosonization higher genus}. 
The set of such boundary conditions provides the {\it spin structure} of the fermionic model on the underlying higher genus Riemann surface.

The function $\mathcal{F}_n(x)$ entering in (\ref{Fn}) is known explicitly only for very few models. 
One of the most important ones is the free  boson compactified on a circle of radius $r_{\rm circle}$.  
In this case,  $\mathcal{F}_n(x)$ is \cite{cct-09}
\begin{equation}
\label{Fnv}
\mathcal{F}_n(x)=
\frac{\Theta\big({\bf 0}|\eta\tau \big)\,\Theta\big({\bf 0}|\tau/\eta\big)}{
[\Theta\big({\bf 0}|\tau\big)]^2} \,,
\end{equation}
where the parameter $\eta= 2r_{\rm circle}^2$ is related to the compactification radius and
$\tau$ is the  $(n-1)\times(n-1)$ period matrix of the Riemann surface $\mathcal{R}_n$, whose elements read \cite{cct-09}
\begin{equation} 
\label{eq:tau}
 \tau_{i,j} \,=\, \iu\,  \frac{2}{n} \sum_{k=1}^{n-1} \sin(\pi k/n) \, \frac{{}_2F_1(k/n,1-k/n;1;1-x)}{{}_2F_1(k/n,1-k/n;1;x)} \, \cos[2\pi (k/n)(i-j)]\,.
\end{equation}
This period matrix can be written by employing the canonical homology basis shown in Figs.\,\ref{fig:cage} and \ref{fig:R4} \cite{cct-09, ctt-14}.
It is worth remarking that, since $x\in (0,1)$, the period matrix $ \tau(x)$ is purely imaginary. 
In (\ref{Fnv}) the Riemann theta function $\Theta$ \cite{theta books,Fay book} is given by
\begin{equation}
\label{theta Riemann def 0}
\Theta(\boldsymbol{z}|M)\,\equiv\,
\sum _{\boldsymbol{m}\,\in\,\mathbb{Z}^{n-1}}
e^{
\,{\rm i} \pi \, \boldsymbol{m}^{\rm t} \cdot M \cdot \boldsymbol{m}
+2\pi {\rm i} \, \boldsymbol{m}^{\rm t}\cdot \boldsymbol{z} 
}\,,
\end{equation}
as function of the $n-1$ dimensional complex vector $\boldsymbol{z}$  and of 
the $(n-1) \times (n-1)$ matrix  $M$, which must be symmetric and with positive imaginary part.

For the CFTs we are interested in, namely the modular invariant Dirac fermion and the critical Ising model, the scaling functions $\mathcal{F}_n (x)$ are also known  \cite{cct-11}.
In particular, for the modular invariant Dirac fermion the  function $\mathcal{F}_n (x)$ reads
\be
\label{Fn Dirac}
\mathcal{F}^{\textrm{Dirac}}_n (x) 
\,=\,
\frac{1}{2^{n-1}} \sum_{\boldsymbol{e}}
(-1)^{4\boldsymbol\varepsilon\cdot\boldsymbol\delta} \,
\bigg| 
\frac{\Theta[\boldsymbol{e}] ({\bf 0}|\tau)}{\Theta ({\bf 0}|\tau)}
\bigg|^2 ,
\ee
and for the Ising model it is given by
\be
\label{Fn Ising}
\mathcal{F}^{\textrm{Ising}}_n (x) 
\,=\,
\frac{1}{2^{n-1}} \sum_{\boldsymbol{e}}
(-1)^{4\boldsymbol\varepsilon\cdot\boldsymbol\delta} \,
\bigg| 
\frac{\Theta[\boldsymbol{e}] ({\bf 0}|\tau)}{\Theta ({\bf 0}|\tau)}
\bigg| \,,
\ee
being the period matrix  $\tau$ is \eqref{eq:tau}.
In these cases $\mathcal{F}_n (x)$ is written in terms of the Riemann theta function with characteristic,  which generalises (\ref{theta Riemann def 0})  as follows \cite{theta books, Fay book}
\be
\label{riemann theta def}
\Theta[\boldsymbol{e}](\boldsymbol{z}|M) \,
\equiv
\sum _{\boldsymbol{m}\,\in\,\mathbb{Z}^{n-1}}
e^{
\,{\rm i} \pi
(\boldsymbol{m}+\boldsymbol{\varepsilon})^{\rm t}
\cdot M \cdot(\boldsymbol{m}+\boldsymbol{\varepsilon})
+2\pi {\rm i} \,
(\boldsymbol{m}+\boldsymbol{\varepsilon})^{\rm t}\cdot (\boldsymbol{z} + \boldsymbol{\delta})
}\,,
\qquad
\boldsymbol{e}  
\equiv 
\bigg( \hspace{-.1cm}
\begin{array}{c}
 \boldsymbol{\varepsilon}\\
 \boldsymbol{\delta}
 \end{array}\hspace{-.1cm}\bigg)\,,
\ee
where $\boldsymbol{z}$ and $M$ are defined as in (\ref{theta Riemann def 0}).
The characteristic $\boldsymbol{e}$ of the Riemann theta function is given by  the  $n-1$ dimensional vectors $\bm{\varepsilon}$ and $\bm{\delta} $, whose entries  are either $0$ or $1/2$.
Notice that (\ref{riemann theta def}) becomes (\ref{theta Riemann def 0}) when $\bm{\varepsilon} = \bm{\delta} = \boldsymbol{0} $.
The characteristic $ \boldsymbol e $ specifies the set of boundary conditions along the $a$ and $b$ cycles, providing the spin structures of the fermionic model \cite{bosonization higher genus}. 
In particular $\boldsymbol{\varepsilon}$ gives the boundary conditions along the $a$ cycles ($\varepsilon_k = 0$ for antiperiodic b.c.\ around $a_k$ and $\varepsilon_k = 1/2$ for periodic b.c.), while $\boldsymbol{\delta}$ is fixed through the boundary conditions along the $b$ cycles ($\delta_k = 0$ for antiperiodic b.c.\ around $b_k$ and $\delta_k = 1/2$ for periodic b.c.).

The expressions \eqref{Fn Dirac} and \eqref{Fn Ising} tell us that the moments of the reduced density matrix can be computed as a sum of fermionic partition functions on $\mathcal R_n$ with all possible choices of fermionic boundary conditions.

As function of $\boldsymbol{z}$, the parity of the Riemann theta function (\ref{riemann theta def}) is given by the parity of the integer number $4\boldsymbol\varepsilon\cdot\boldsymbol\delta$ (indeed $\Theta[\boldsymbol{e}](-\boldsymbol{z}|M) = (-1)^{4\boldsymbol\varepsilon\cdot\boldsymbol\delta} \, \Theta[\boldsymbol{e}](\boldsymbol{z}|M) $),
which is, by definition, also the parity of the characteristic $\boldsymbol{e}$.
Among the $2^{2(n-1)}$ possible characteristics, $2^{n-2}(2^{n-1}+1)$ are even and $2^{n-2}(2^{n-1}-1)$ are odd. 
Thus, in the expressions for $\mathcal{F}_n(x)$ given above, where $\boldsymbol{z}=\boldsymbol{0}$, only the Riemann theta functions with even characteristics are non vanishing. 
This implies that the terms with odd characteristics can be multiplied  by arbitrary functions. 
In (\ref{Fn Dirac}) and (\ref{Fn Ising}) we have introduced a minus sign in front of all the terms with odd characteristics because this choice facilitates the identification of their lattice counterparts, as it will be discussed in \S\ref{sec:renyi}.

The formulas (\ref{Fn Dirac}) and (\ref{Fn Ising}) have been found by employing old results about the bosonization on higher genus Riemann surfaces \cite{bosonization higher genus}. 
In particular, we remark that $\mathcal{F}^{\textrm{Dirac}}_n (x) $ in (\ref{Fn Dirac}) comes from (\ref{Fnv}) for a specific value of the compactification radius ($\eta=1/2$ in the notation of \cite{cct-09}).

Plugging (\ref{Fn Dirac}) and (\ref{Fn Ising}) into (\ref{Fn}) with the proper choice of the central charge $c$ and the coefficient $c_n$, one finds the moments of the reduced density matrix for the modular invariant Dirac fermion and for the Ising model respectively. 
We find it convenient to split the various terms occurring in the resulting sums by introducing the following notation\footnote[4]{In order to enlighten some forthcoming expressions, in (\ref{Jn Omega def}) and (\ref{tilde Jn Omega def}) we have slightly changed the notation for $\Omega_n$ and $\widetilde{\Omega}_n$ with respect to the one adopted in \cite{ctc-15-2}.}
\be
\label{Jn Omega def}
J_n
\equiv 
 \frac{ c_n^2}{\big[\ell_1\ell_2(1-x)\big]^{2\Delta_n}} \,,
\qquad
\Omega_n[ 2\boldsymbol e ]  
\equiv 
 \bigg|\frac{\Theta[ \boldsymbol e ] 
({\bf 0}|\tau(x))}{\Theta({\bf 0}|\tau(x))}\bigg| \,.
\ee
In terms of these expressions, $\Tr \rho_A^n$ for the Dirac fermion and the Ising model can be written respectively as 
$\Tr \rho_A^n  =  \tfrac{1}{2^{n-1}} \, J_n^{\textrm{Dirac}} \sum_{\boldsymbol e} \Omega_n[ 2\boldsymbol e ]^2  $ and 
$\Tr \rho_A^n  =  \tfrac{1}{2^{n-1}} \, J_n^{\textrm{Ising}} \sum_{\boldsymbol e} \Omega_n[ 2\boldsymbol e ]  $, where $J_n$ is different in the two models because of the central charge and the non-universal constant $c_n$.

Performing the replica limit for these expressions in order to get the entanglement entropy $S_A$ or the mutual information is still an open problem (see \cite{dct-15} for numerical extrapolations).

The moments of the partial transpose $\Tr (\rho_A^{T_2})^n $ in a conformal field theory can be also computed as the four point function of twist fields given by $\langle {\cal T}_n(u_1)\bar {\cal T}_n(v_1) \bar {\cal T}_n(u_2) {\cal T}_n(v_2) \rangle$ \cite{cct-neg-letter}, where $u_1<v_1<u_2<v_2$ are the endpoints of the disjoint intervals.
Thus, they admit the following universal scaling form 
\be
\label{tr rhoAn T2 cft}
\Tr \big(\rho_A^{T_2}\big)^n 
\,=\,
  c_n^2 \left( \frac{(u_2-u_1)(v_2-v_1)}{(v_1-u_1)(v_2-u_2)(v_2-u_1)(u_2-v_1)} \right)^{2\Delta_n}  
  \mathcal{G}_{n}(x) \,,
\ee
being $c_n$ the non-universal constant defined in (\ref{Renyi:asymp}),
$\mathcal{G}_{n}(x)$ a new universal scaling function and $x$ the cross ratio (\ref{cross ratio def}).
The scaling functions ${\cal F}_n(x)$ and ${\cal G}_n(x)$, introduced in (\ref{Fn}) and (\ref{tr rhoAn T2 cft}) respectively, are related as follows \cite{cct-neg-letter,cct-neg-long}
\be
\label{G via F}
{\cal G}_n(x) = (1-x)^{4\Delta_n} \, \mathcal{F}_{n}\bigg(\frac{x}{x-1}\bigg) \,,
\ee
where it is worth remarking that $x/(x-1) \in (-\infty, 0)$ when $x\in (0,1)$.
This means that the function ${\cal F}_n(x)$ defined for $x\in (0,1)$ in (\ref{Fn}) has to be properly extended to a function ${\cal F}_n(x, \bar{x})$ defined on the whole complex plane.
Then, its restriction to the real negative axis provides the function  ${\cal G}_n(x)$ defined for $x\in (0,1)$ according to (\ref{G via F}).
Notice that in the ratio $\Tr (\rho_A^{T_2})^n / \Tr \rho_A^n = (1-x)^{4\Delta_n} \mathcal{F}_{n}(\tfrac{x}{x-1})/ \mathcal{F}_{n}(x)$ the non universal constants $c_n$ and the dimensionfull factors simplify, leaving only a universal scale invariant quantity.

\begin{figure}
\vspace{.4cm}
\begin{center}
\includegraphics[width=1\textwidth]{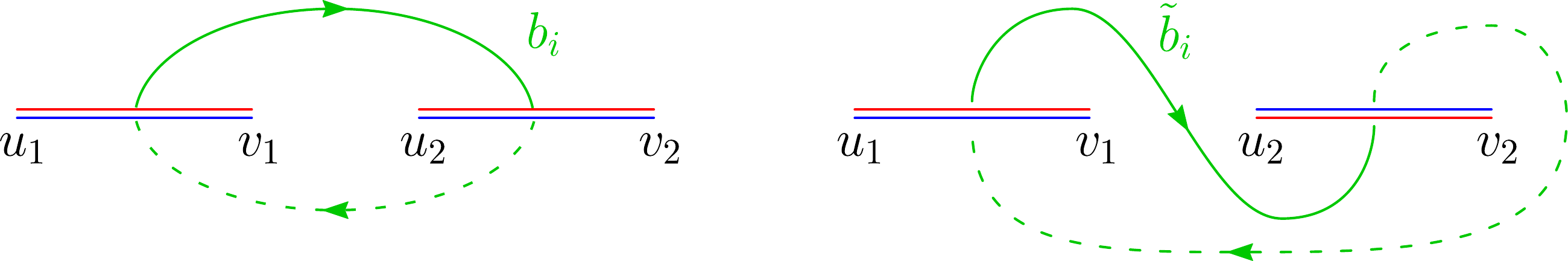}
\end{center}
\vspace{-.1cm}
\caption{The $i$-th $b$ cycle for the Riemann surfaces $\mathcal{R}_n$ (left) and $\widetilde{\mathcal{R}}_n$ (right): the solid part of the curve belongs to the $i$-th sheet and the dashed part to the $(i+1)$-th sheet. 
As for the $a$ cycles, they coincide for both $\mathcal{R}_n$ and $\widetilde{\mathcal{R}}_n$ and in Fig.\,\ref{fig:R4} they are shown explicitly for $n=4$.
}
\label{fig:bcycles}
\end{figure}

Since  (\ref{tr rhoAn T2 cft}) is obtained by joining $n$ replicas of the model in a proper way \cite{cct-neg-letter,cct-neg-long}, also the  $n$-th moment $\Tr (\rho_A^{T_2})^n $ can be evaluated as the partition function of the model on a particular Riemann surface $\widetilde{\mathcal{R}}_n$.
Such Riemann surface has genus $n-1$ and it is genuinely different from $\mathcal{R}_n$  when $n>2$.
Instead, $\widetilde{\mathcal{R}}_2$ and $\mathcal{R}_2$ are the same torus because their moduli are related by a modular transformation \cite{cct-neg-long}. 

The Riemann surface  $\widetilde{\mathcal{R}}_n$ is defined by the complex constraint $y^n = (z-u_1)(z-v_2) [(z-v_1)(z-u_2)]^{n-1}$ in $\mathbb{C}^2$, which has been obtained by exchanging $u_2 \leftrightarrow v_2$ in the equation defining $\mathcal{R}_n$.
The surface $\widetilde{\mathcal{R}}_n$ can be constructed by joining properly the $n$ replicas \cite{cct-neg-letter, cct-neg-long} and this procedure is different from the one employed to define $\mathcal{R}_n$.
In particular, since we are considering the partial transpose with respect to $A_2$, the edges of the slits along the $n$ copies of $A_1$ are attached cyclically like in $\mathcal{R}_n$, but the edges along $A_2$ are attached in the opposite way: if along $A_1$ the upper edge in the $i$-th sheet is identified with the lower edge in the $(i+1)$-th sheet, along $A_2$ the lower edge in the $i$-th sheet should be merged  with the upper edge of $A_2$ in the $(i+1)$-th sheet. We refer to Fig.\,4 of \cite{cct-neg-long} for an explicit representation.

As for the canonical homology basis $\{\tilde{a}_r, \tilde{b}_r \,; \, 1\leqslant r\leqslant n-1\}$ for $\widetilde{\mathcal{R}}_n$, since $\widetilde{\mathcal{R}}_n$ and $\mathcal{R}_n$ differ only 
 for the joining of the sheets along $A_2$, it is natural to choose $\tilde{a}_r = a_r$ (see Fig.\,\ref{fig:R4}), while the generic cycle $\tilde{b}_r $ is shown in the right panel of Fig.\,\ref{fig:bcycles}.
 This basis has been already employed in \cite{cct-neg-long, ctc-15-2} and the period matrix $\tilde{\tau}(x)$ of $\widetilde{\mathcal{R}}_n$ reads
\begin{equation} 
\label{eq:Q}
\tilde{\tau}(x) = \tau\big(x/(x-1)\big) = \mathcal{R} + \iu \,\mathcal{I}  \,,
\qquad 
\mathcal{R} = \frac{1}{2}\, \mathcal{Q} \,,
\end{equation}
where the generic element of $\tau(x)$ is given by (\ref{eq:tau}).
The symmetric matrices $\mathcal{R}$ and $\mathcal{I}$ are the real and imaginary part of $\tilde{\tau}$ respectively. 
It is worth remarking that $\mathcal{R}$ is independent of $x$ and its form is particularly simple: it is an integer symmetric matrix $\mathcal{Q}$ multiplied by $1/2$, where 
$\mathcal{Q}$ is a tridiagonal matrix having $2$'s along the principal diagonal and $-1$'s along the secondary ones \cite{ctc-15-2}.

We find it convenient to write explicitly the functions $\mathcal{F}_{n}(\tfrac{x}{x-1})$ entering in (\ref{G via F}) for the CFT models we are dealing with. 
For the modular invariant Dirac fermion it reads
\be
\label{tilde Fn Dirac}
\mathcal{F}^{\textrm{Dirac}}_n \bigg(\frac{x}{x-1}\bigg)
=\,
\frac{1}{2^{n-1}} \sum_{\boldsymbol{e}}
(-1)^{4\boldsymbol\varepsilon\cdot\boldsymbol\delta} \,
\bigg| 
\frac{\Theta[\boldsymbol{e}] ({\bf 0}|\tilde{\tau})}{\Theta ({\bf 0}|\tilde{\tau})}
\bigg|^2,
\ee
and for the Ising model it is given by 
\be
\label{tilde Fn Ising}
\mathcal{F}^{\textrm{Ising}}_n \bigg(\frac{x}{x-1}\bigg)
=\,
\frac{1}{2^{n-1}} \sum_{\boldsymbol{e}}
(-1)^{4\boldsymbol\varepsilon\cdot\boldsymbol\delta} \,
\bigg| 
\frac{\Theta[\boldsymbol{e}] ({\bf 0}|\tilde{\tau})}{\Theta ({\bf 0}|\tilde{\tau})}
\bigg| \,,
\ee

Let us stress again that the minus sign in front of the terms with odd characteristics (which are identically zero) has been introduced to simplify 
the identification of the lattice quantities which vanish in the scaling limit (see \S\ref{sec:moments pt2}). 

Plugging (\ref{tilde Fn Dirac}) and (\ref{tilde Fn Ising}) into (\ref{G via F}) first and then into (\ref{tr rhoAn T2 cft}) with the corresponding central charges and coefficients $c_n$, one obtains the moments $\Tr (\rho_A^{T_2})^n $ for the modular invariant Dirac fermion and the Ising model. 
We find it convenient to introduce expressions similar to (\ref{Jn Omega def}) as follows
\be
\label{tilde Jn Omega def}
\tilde{J}_n
\equiv 
 c_n^2 \left( \frac{1-x}{\ell_1\ell_2} \right)^{2\Delta_n},
\qquad
\widetilde{\Omega}_n[ 2\boldsymbol e ]  
\equiv 
 \bigg|\frac{\Theta[ \boldsymbol e ] 
({\bf 0}|\tilde\tau(x))}{\Theta({\bf 0}|\tilde\tau(x))}\bigg| \,.
\ee
By employing (\ref{tilde Jn Omega def}), the moments of the partial transpose for the Dirac fermion and the Ising model become respectively 
$\Tr (\rho_A^{T_2})^n  =  \tfrac{1}{2^{n-1}} \, \tilde{J}_n^{\textrm{Dirac}} \sum_{\boldsymbol e} \widetilde{\Omega}_n[ 2\boldsymbol e ]^2  $  and 
$\Tr (\rho_A^{T_2})^n   =  \tfrac{1}{2^{n-1}} \, \tilde{J}_n^{\textrm{Ising}} \sum_{\boldsymbol e} \widetilde{\Omega}_n[ 2\boldsymbol e ]  $, where $\tilde{J}_n$ is model dependent as explained above for $J_n$.

The replica limit (\ref{neg replica limit}) giving the logarithmic negativity from these expressions of the moments is still an open problem.
Even numerical extrapolations like the ones performed successfully in \cite{dct-15} to get the mutual information are problematic for the replica limit (\ref{neg replica limit}) because, since only the sequence of even integers is required, one needs higher values of $n$ to get a stable extrapolation and computing the Riemann theta functions is computationally hard for high orders.

\subsection{The dihedral symmetry of the Riemann surfaces}
\label{sec:dihedral symmetry riemann surf}

The dihedral symmetry $\mathbb{Z}_n \times \mathbb{Z}_2$ has been  discussed in \cite{hlr-13, ctt-14} for $\mathcal{R}_n$ and in  \cite{ctc-15-2} for $\widetilde{\mathcal{R}}_n$.

The $\mathbb{Z}_n$ symmetry comes from the fact that both $\mathcal{R}_n$  and $\widetilde{\mathcal{R}}_n$ have been obtained through a replica construction, and therefore they enjoy the $\mathbb{Z}_n$ invariance under the cyclic permutation of the $n$ sheets.
Instead, the $\mathbb{Z}_2$ symmetry originates from the fact that the endpoints of the intervals are located along the real axis.
Indeed, since the complex equations defining $\mathcal{R}_n$ and $\widetilde{\mathcal{R}}_n$ are invariant under complex conjugation,  both these Riemann surfaces remain invariant by taking the sheets in the reversed order and reflecting all of them with respect to the real axis.

The modular transformations of a genus $g$ Riemann surface can be identified with the group of the integer symplectic matrices $Sp(2g,\mathbb{Z})$ \cite{bosonization higher genus}.
These transformations act on the period matrix and reshuffle the characteristics.
In particular, they do not change the parity of a given  characteristic.
The transformations of the dihedral symmetry can be identified with a subgroup of the symplectic matrices which leave the functions in (\ref{Jn Omega def}) and (\ref{tilde Jn Omega def}) 
separately invariant. 
The symplectic matrices implementing the dihedral symmetry have been written explicitly in \cite{hlr-13, ctt-14} for $\mathcal{R}_n$ and in \cite{ctc-15-2} for $\widetilde{\mathcal{R}}_n$ (notice that the matrices for the $\mathbb{Z}_2$ symmetry  are different in the two surfaces).
Thus, beside the vanishing terms with odd characteristics in the sums (\ref{Fn Dirac}), (\ref{Fn Ising}), (\ref{tilde Fn Dirac}) and  (\ref{tilde Fn Ising}), the dihedral symmetry leads to further degeneracies among the non vanishing terms. 
Indeed, the terms whose even characteristics are related by one of the modular transformations associated to the dihedral symmetry are equal.
This implies that the sums for the moments can be written in a simpler form by choosing a representative term for each equivalence class, whose coefficient is the cardinality of the corresponding equivalence class.
By implementing the dihedral symmetry in (\ref{Fn Dirac}), (\ref{Fn Ising}), (\ref{tilde Fn Dirac}) and  (\ref{tilde Fn Ising}), one can slightly reduce  the exponentially large (in $n$) number of terms occurring in these sums.

We find it instructive to write explicitly $\mathcal F_n(x) $ for the Ising model in the simplest cases of $n=2$ and $n=3$. By specialising (\ref{Fn Ising}) to these values of $n$ and implementing the dihedral symmetry discussed above, the results are respectively
\bea
\label{Fn ising n=2}
& & \hspace{-2cm}
\mathcal F_2^\text{Ising}(x) =
\frac{1}{2}\, \bigg(
1
+\Omega_2\bigg[\hspace{-.1cm} \begin{array}{c}   0 \\ 1   \end{array} \hspace{-.1cm} \bigg]
+\Omega_2\bigg[\hspace{-.1cm} \begin{array}{c}   1 \\ 0   \end{array} \hspace{-.1cm} \bigg]
-\Omega_2\bigg[\hspace{-.1cm} \begin{array}{c}   1 \\ 1   \end{array} \hspace{-.1cm} \bigg]
\,\bigg) \,,
\\
\rule{0pt}{.8cm}
\label{Fn ising n=3}
& & \hspace{-2cm}
\mathcal F_3^\text{Ising}(x) =
\frac{1}{4}\, \bigg(
1
+3\,\Omega_3\bigg[\hspace{-.1cm} \begin{array}{cc}  0 & 0 \\ 0 & 1  \end{array} \hspace{-.1cm} \bigg]
+3\,\Omega_3\bigg[\hspace{-.1cm} \begin{array}{cc}  0 & 1 \\ 0 & 0  \end{array}\hspace{-.1cm} \bigg]
+3\,\Omega_3\bigg[\hspace{-.1cm} \begin{array}{cc}  0 & 1 \\ 1 & 0  \end{array}\hspace{-.1cm} \bigg]
-6\,\Omega_3\bigg[\hspace{-.1cm} \begin{array}{cc}  0 & 1 \\ 1 & 1  \end{array}\hspace{-.1cm} \bigg]
\,\bigg) \,,
\eea
where we have written also the vanishing terms with odd characteristics, which occur with a minus sign \cite{ginsparg}.
The corresponding expressions for the Dirac model can be easily obtained from (\ref{Fn Dirac}), or by simply replacing each $\Omega_n[ 2\boldsymbol e ] $ in (\ref{Fn ising n=2}) and (\ref{Fn ising n=3}) with $\Omega_n[ 2\boldsymbol e ]^2 $.

As for the moments of the partial transpose (\ref{tr rhoAn T2 cft}) and (\ref{G via F}) for the Ising model, by specialising (\ref{tilde Fn Ising}) to $n=2$ and $n=3$ and implementing the dihedral symmetry we find
\bea
\label{Fn tilde ising n=2}
& & \hspace{-2.4cm}
\mathcal F_2^\text{Ising}\left(\frac{x}{x-1}\right) =
\frac{1}{2}\, \bigg(
1
+\widetilde\Omega_2\bigg[\hspace{-.1cm} \begin{array}{c}   0 \\ 1   \end{array}\hspace{-.1cm} \bigg]
+\widetilde\Omega_2\bigg[\hspace{-.1cm} \begin{array}{c}   1 \\ 0   \end{array}\hspace{-.1cm} \bigg]
-\widetilde\Omega_2\bigg[\hspace{-.1cm} \begin{array}{c}   1 \\ 1   \end{array}\hspace{-.1cm} \bigg]
\,\bigg) \,,
\\
\label{Fn tilde ising n=3}
\rule{0pt}{.8cm}
& & \hspace{-2.4cm}
\mathcal F_3^\text{Ising}\left(\frac{x}{x-1}\right) =
\frac{1}{4}\, \bigg(
1
+3\,\widetilde\Omega_3\bigg[\hspace{-.1cm} \begin{array}{cc}  0 & 0 \\ 0 & 1  \end{array}\hspace{-.1cm} \bigg]
+6\,\widetilde\Omega_3\bigg[\hspace{-.1cm} \begin{array}{cc}  0 & 1 \\ 1 & 0  \end{array}\hspace{-.1cm} \bigg]
-3\,\widetilde\Omega_3\bigg[\hspace{-.1cm} \begin{array}{cc}  0 & 1 \\ 0 & 1  \end{array}\hspace{-.1cm} \bigg]
-3\,\widetilde\Omega_3\bigg[\hspace{-.1cm} \begin{array}{cc}  0 & 1 \\ 1 & 1  \end{array}\hspace{-.1cm} \bigg]
\,\bigg) .
\eea
As above, the corresponding expressions for the Dirac fermion can be written straightforwardly by substititung each $\Omega_n[ 2\boldsymbol e ] $ in (\ref{Fn tilde ising n=2}) and (\ref{Fn tilde ising n=3}) with $\Omega_n[ 2\boldsymbol e ]^2 $.

%%%%%%%%%%%%%%%%%%%%%%%%%%%%%%%%%%%%%%%%%%%%

\section{Spin structures in the moments of the reduced density matrix and its partial transpose}
\label{sec:scaling reg}

In this section we employ the coherent states formalism to identify the analytic expressions giving the scaling limit of the terms occurring in the sums for 
$\Tr \rho_A^n$ and $\Tr( \rho^{T_2}_A)^n$ on the lattice discussed in \S\ref{sec:rev lattice}.
These are the partition functions of the corresponding fermionic model on the underlying Riemann surface with a specific spin structure.

\subsection{Reduced density matrix}
\label{sec:rdm}

In this subsection we provide a representation of the reduced density matrix on the lattice in terms of the fermionic coherent states \cite{bravyi, kleinert}.
The generalisation to a continuum spatial dimension is straightforward.

In the fermionic representation let us consider the local Hilbert space of the $j$-th fermion, whose basis is given by the two vectors
$\ket{1}$ and $\ket{0}$, telling whether the fermion occurs or not respectively.
The operators $c_j$ and $c_j^\dag$ act as the creation and annihilation operators on these vectors, namely $c_j\ket{0}=c_j^\dagger \ket{1}=0$, while $c_j^\dagger \ket{0}= \ket{1}$ and $c_j \ket{1}= \ket{0}$.

Given the Majorana operators (\ref{a from c def}), let us define the following single site operators 
\begin{equation}
\label{majorana op}
 a_j^x = \Sigma_j^x\,,   \qquad a_j^y = -\Sigma_j^y \,,  \qquad \iu \,a_j^xa_j^y = \Sigma_j^z \,,
\end{equation}
which satisfy the algebra of the Pauli matrices.
Nevertheless, it is worth remarking that, since they anticommute at different sites, they are not spin operators and should not be 
confused with the $\sigma_j^\alpha$ in (\ref{eq:HXX Hising}). 

For the $j$-th site, it is useful to introduce also the following unitary operator 
\begin{equation}
\label{Uk def}
 U_{\alpha}^{(k)} = e^{\iu\frac{\alpha}{2}\Sigma_j^k} = \cos\left(\alpha/2\right)\mathbb{I} + \iu\sin\left(\alpha/2\right) \Sigma_j^k \,,
\end{equation}
whose action on the operators in (\ref{majorana op}) can be obtained from the following relation
\begin{equation}
 U_{-\alpha}^{(k)}\, \Sigma_j^{b}\, U_{\alpha}^{(k)} 
 = \big[ \delta_{kb} + (1-\delta_{kb}\cos\alpha) \big]\Sigma_j^b +  (\sin\alpha) \varepsilon_{kb\ell} \,\Sigma_j^\ell \,,
\end{equation}
where $\varepsilon_{kb\ell} $ is the totally antisymmetric tensor such that $\varepsilon_{xyz} =1$.

Considering a real Grassmann variable $\theta$, since $\theta^2 =0$, a generic function of such variable can be written as $ f(\theta) = f_0 + f_1\theta$, where $f_i$ are real or complex numbers.
Given two real Grassmann variables $\theta_1$ and $\theta_2$, we have that $\theta_i^2=0$ for $i\in \{1,2\}$ and $\theta_1\theta_2 = -\theta_2\theta_1$.
A complex Grassmann variable $\zeta$ can be built from two real Grassmann variables as $ \zeta = (\theta_1+\iu\,\theta_2)/\sqrt{2}$ and its complex conjugate reads $ \conj\zeta = (\theta_1-\iu\,\theta_2)/\sqrt{2}$.
Integrating over a complex Grassmann variable is like taking the derivative; indeed
\begin{equation}
 \int\df\zeta =
 \int\df\zeta \,\zeta = 
 \int\df\zeta \,\conj\zeta = 0 \,,
 \qquad
 \int\df\zeta \,\zeta\,\conj\zeta = 1 \,.
\end{equation}
The coherent states for a single site are defined as follows
\begin{equation}
\label{coherent states}
 \ket{\zeta} = \ket{0} - \zeta\ket{1} \,,
 \qquad 
 \bra{\zeta} = \bra{0} + \conj\zeta\bra{1} \,,
\end{equation}
where $\ket{0}$ e $\ket{1}$ have been introduced above.
The peculiar property of these states is that $c\ket{\zeta} = \zeta\ket{\zeta} $ and $\bra{\zeta}c^\dag = \bra{\zeta} \conj{\zeta}$,
as one can easily check by employing that $\zeta$ commutes with $\ket{0}$ and anticommutes with $c$, $c^\dagger$ and $\ket{1}$.
The coherent states in (\ref{coherent states}) do not form an orthonormal basis. 
A completeness relation and a formula for the trace of an operator $O$ are given respectively by
\begin{equation} 
\label{eq:identity and trace}
 \mathbb{I} = \intf{\zeta}\,e^{-\conj\zeta\zeta}\ket{\zeta}\bra{\zeta} \,,
 \qquad 
 \textrm{Tr}\, O = \intf{\zeta}\,e^{-\conj\zeta\zeta}\bra{-\zeta} O\ket{\zeta} \,.
\end{equation}
In the following we will also need
\be
\label{eta zeta scalar product}
\braket{\zeta|\eta} = 1+\conj\zeta\eta = \, e^{\zeta^\ast \eta} \,.
\ee
Another useful property to remark is $\iu a^x a^y\ket{\zeta} = \ket{-\zeta}$, which can be easily derived from (\ref{a from c def}), (\ref{coherent states}) and the above definition of the operators $c$ and $c^\dagger$.

A coherent state for the whole lattice is constructed by simply taking the tensor product of the single site coherent states just discussed, i.e. $\ket{\zeta(x)} = \otimes_i \ket{\zeta}_i $  (here we have restored the lattice index and $x$ is a discrete variable labelling the chain). 
When the whole system is in the ground state, the density matrix is  $\rho = \ket{\Psi}\bra{\Psi}$ and its matrix element with respect to two generic coherent states reads
\begin{equation}
\label{rho mat element}
 \rho(\zeta(x),\eta(x)) \,=\,  e^{-\conj\zeta \eta} \braket{\zeta(x)|\Psi}\braket{\Psi|\eta(x)},
\end{equation}
where $\conj\zeta \eta= \sum_x \conj\zeta(x) \, \eta(x)$ and $e^{-\conj\zeta \eta} $ is the normalization factor obtained through (\ref{eta zeta scalar product}).
It is worth remarking that the above formulas are given for a discrete system, but they can be easily adapted to a continuum system.

\begin{figure}
\vspace{.3cm}
\includegraphics[width=1.\textwidth]{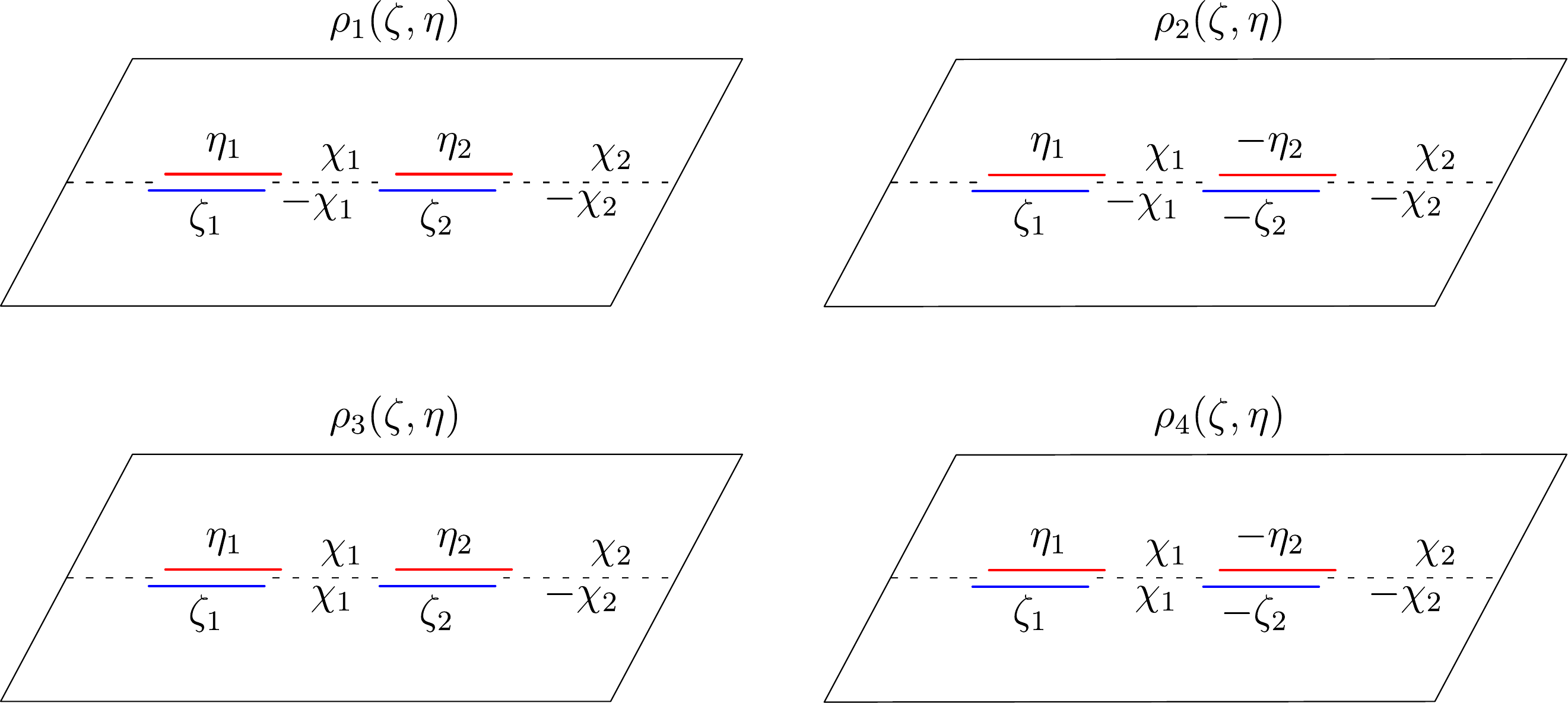}
\vspace{-.3cm}
\caption{Path integral representation in the coherent state basis of the terms occurring in $\Tr\rho_A^n$ (see \S\ref{sec:rdm} and \S\ref{sec:renyi}).
} 
\label{fig:shee}
\end{figure} 

The reduced density matrix of $A=A_1 \cup A_2$ is obtained by separating the degrees of freedom in $A$ and the ones in $B$ first and then tracing over the latter ones. 
Denoting by $\ket{\zeta_A(x_A)}$ and $\ket{\zeta_B(x_B)}$ the coherent states on $A$ and $B$ respectively, the coherent state on the full system can be written as $ \ket{\zeta(x)} =  \ket{\zeta_A(x_A)}\otimes \ket{\zeta_B(x_B)} = \ket{\zeta_A(x_A),\,\zeta_B(x_B)}$, where in the last step we have introduced the notation that will be adopted hereafter.

In \S\ref{subsec rho_A} it has been discussed the spin reduced density matrix $\rho_A$, finding that its $n$-th moment $\Tr \rho_A^n$ can be computed by taking the trace over the fermionic degrees of freedom of a combination made by the four Gaussian fermionic operators (\ref{rho 1234 def}).
In the following we study matrix elements of these four matrices with respect to two generic coherent states $\ket{\zeta_A(x_A)}$ and $\ket{\eta_A(x_A)}$.

Considering the density matrix $\rho_A^{\,{\bf 1}} $ first, its matrix element reads
\be
\label{eq:rho1 path integral}
 \rho_1(\zeta_A,\eta_A) \,= \,
 e^{-\conj\zeta_A \eta_A} 
 \int D\chi_B^\ast \, D\chi_B
 \, e^{-\conj\chi_B \,\chi_B} \braket{\zeta_{A_1}, \zeta_{A_2}; -\chi_{B_1}, -\chi_{B_2}|\Psi}  \braket{\Psi|\eta_{A_1} ,  \eta_{A_2};\,\chi_{B_1},\chi_{B_2}} ,
\ee
where $\zeta_A = \zeta_A(x_A)$, $\eta_A = \eta_A(x_A)$ and, for later convenience, within the matrix elements the contributions of the various blocks have been separated to highlight the boundary conditions and the joining conditions.
In the exponent occurring in (\ref{eq:rho1 path integral}) we have $\conj\chi_B \,\chi_B = \conj\chi_{B_1} \,\chi_{B_1} + \conj\chi_{B_2} \,\chi_{B_2}$.
As for the integration measure, it is given by  $D \chi_B^\ast \, D\chi_B =   \prod_{x\in B} {\rm d}\chi_B^\ast(x) \, {\rm d}\chi_B(x)$,  which can also be written as $D \chi_B^\ast \, D\chi_B = \prod_{x \in B_1} {\rm d}\chi_B^\ast(x) \, {\rm d}\chi_B(x) \prod_{x\in B_2} {\rm d}\chi_B^\ast(x) \, {\rm d}\chi_B(x)$.
The minus sign within the matrix element in (\ref{eq:rho1 path integral}) comes from the trace over $B$, according to (\ref{eq:identity and trace}). 

The formula (\ref{eq:rho1 path integral}) is given for a lattice system but it can be easily generalised to a continuum spatial dimension by interpreting the discrete product in the integration measure as a path integral along the spatial direction and the discrete sum $\conj\zeta_A \eta_A$ as an integral over $A$.
The braket $\braket{\Psi|\eta_{A_1} ,  \eta_{A_2};\,\chi_{B_1},\chi_{B_2}}$ in the continuum corresponds to the fermionic path integral on the upper half plane where the boundary conditions $\eta_{A_j}$ and $\chi_{B_j}$ (with $j=1,2$) are imposed in $A_j$ and $B_j$ respectively, just above the real axis.
In a similar way, $\braket{\zeta_{A_1}, \zeta_{A_2}; -\chi_{B_1}, -\chi_{B_2}|\Psi} $ is a fermionic path integral on the lower half plane with the proper boundary conditions explicitly indicated.
Performing the  trace over $B$  corresponds to set the fields along $B$ equal (but with opposite sign) and summing over all the possible field configurations.
The net result is a path integral over the whole plane with two open slits along $A_1$ and $A_2$, where the boundary conditions $\eta_A$ and $\zeta_A$ are imposed along the lower and the upper edges of $A$ respectively. This is represented pictorially in the top left panel of Fig.\,\ref{fig:shee}.

The matrix element of $\rho_2$ in (\ref{rho 1234 def}) can be easily obtained once the role of $P_{A_2}$ is understood.
From the observation below (\ref{eta zeta scalar product}) and since $P_{A_2}$ occurs on both sides of $\rho_1$, it is not difficult to realise that the net effect of $P_{A_2}$ is the change of sign for the fermion both above and below the cut along $A_2$, namely
\bea
\label{eq:rho2 path integral}
& &  \hspace{-2.2cm}
 \rho_2(\zeta_A,\eta_A) \,= \, 
  e^{-\conj\zeta_A \eta_A}   \bra{\zeta_{A_1}, \zeta_{A_2} } P_{A_2}   \rho_1 P_{A_2} \ket{\eta_{A_1} ,  \eta_{A_2}}
  \, =\,
 \rho_1(\zeta_{A_1},-\zeta_{A_2};\eta_{A_1},-\eta_{A_2}) 
 \\
 & & \hspace{-1.5cm}
 =\,
 e^{-\conj\zeta_A \eta_A} 
 \int D\chi_B^\ast \, D\chi_B
 \, e^{-\conj\chi_B \,\chi_B} \braket{\zeta_{A_1}, - \zeta_{A_2}; -\chi_{B_1}, -\chi_{B_2}|\Psi}  \braket{\Psi|\eta_{A_1} ,  - \eta_{A_2};\,\chi_{B_1},\chi_{B_2}} .
 \nonumber
\eea
As above, also this expression can be easily adapted to the case of a continuum spatial dimension (see the top right panel of Fig.\,\ref{fig:shee}).

As for the term $\rho_3= \Tr_B (P_{B_1}  |\Psi \rangle \langle \Psi |)$ in  (\ref{rho 1234 def}), the role of $P_{B_1}$ within the trace over $B$ is crucial.
Since for every single site we have $\iu a^x a^y\ket{\zeta} = \ket{-\zeta}$, 
the effect of $P_{B_1}$ is implemented during the integration along $B$ by taking the field above and below $B_1$ with the same sign (while keeping a relative minus sign above and below $B_2$), namely
\begin{equation}
\label{eq:rho3 path integral}
 \rho_3(\zeta_A,\eta_A)\,= \,
 e^{-\conj\zeta_A \eta_A} 
 \int D\chi_B^\ast \, D\chi_B
 \, e^{-\conj\chi_B \,\chi_B} \braket{\zeta_{A_1}, \zeta_{A_2}; \chi_{B_1}, -\chi_{B_2}|\Psi}  \braket{\Psi|\eta_{A_1} ,  \eta_{A_2};\,\chi_{B_1},\chi_{B_2}} .
\end{equation}
It is useful to compare this matrix element with the corresponding one for $\rho_1$ in \eqref{eq:rho1 path integral}.

Finally, the matrix element $\rho_4(\zeta_A,\eta_A)$ of $\rho_4$ in (\ref{rho 1234 def}) is obtained by applying to $\rho_3(\zeta_A,\eta_A)$ the same considerations about the effect of $P_{A_2}$ done to get $\rho_2(\zeta_A,\eta_A)$ from $\rho_1(\zeta_A,\eta_A)$.
The result reads
\bea
\label{eq:rho4 path integral}
& &  \hspace{-2.3cm}
 \rho_4(\zeta_A,\eta_A) 
 \,= \, 
e^{-\conj\zeta_A \eta_A}   \bra{\zeta_{A_1}, \zeta_{A_2} } P_{A_2} \rho_3 P_{A_2} \ket{\eta_{A_1} ,  \eta_{A_2}}
  \, =\, 
 \rho_3(\zeta_{A_1},-\zeta_{A_2};\eta_{A_1},-\eta_{A_2}) 
 \\
 & & \hspace{-.5cm}
=\,    \int D\chi_B^\ast \, D\chi_B
 \, e^{-\conj\chi_B \,\chi_B} \braket{\zeta_{A_1},-\zeta_{A_2};\,\chi_{B_1},\,-\chi_{B_2}|\Psi}  \braket{\Psi|\eta_{A_1},-\eta_{A_2};\,\chi_{B_1},\,\chi_{B_2}} ,
 \nonumber
\eea
The matrix elements $\rho_j(\zeta_A,\eta_A) $ discussed above in the continuum have been represented pictorially in Fig.\,\ref{fig:shee}.

In the continuum limit, the operator $P_C$ in (\ref{P_C def}) should be identified with the following operator
\begin{equation}
\label{P_C cont limit}
 P_{C}=  (-1)^{\int_{C}\text d x\, \bar\psi(x) \psi(x)}  \equiv (-1)^{F_C} ,
\end{equation}
where $F_C$ is the fermionic number operator in the interval $C$ \cite{ginsparg}.
The operator (\ref{P_C cont limit}) is placed along the interval $C$ and it changes the fermionic boundary conditions (from antiperiodic to periodic or viceversa) 
on any cycle crossing the curve $C$.

In  $\rho_2$ and $\rho_4$ (see (\ref{rho 1234 def}))  $P_{A_2}$ occurs both before and after $\rho_A^{\,{\bf 1}} $ and $\rho_A^{B_1}$ respectively.
This means that, in their continuum limit, the  operators $(-1)^{F_{A_2}}$ must be inserted both along the upper edge and the lower edge of $A_2$.
As for $\rho_3$ and $\rho_4$, which are defined in (\ref{rho 1234 def}), the crucial difference with respect to $\rho_1$ and $\rho_2$ is that $P_{B_1}$ occurs through (\ref{fake rhoA});
therefore their continuum limits contain also the operator $(-1)^{F_{B_1}}$ applied once along $B_1$.
Thus, for instance, in the path integral representation of $ \rho_4(\zeta_A,\eta_A) $ the operator $(-1)^F$ occurs both around $A_2$ and along $B_1$.

\subsection{Moments of the reduced density matrix}
\label{sec:renyi}

In this subsection we discuss the scaling limit of the lattice quantities defined by the terms of the sum \eqref{renyi T_n}, finding that 
the scaling limit of the term characterised by the vector $\boldsymbol q$ in such sum is the term associated to a particular spin structure 
characterised by $\boldsymbol e$  in the sums  \eqref{Fn Dirac} and \eqref{Fn Ising}.

In \S\ref{subsec rho_A} we have seen that the $n$-th moment of the reduced density matrix is given by \eqref{renyi T_n}, which contains $2^{2(n-1)}$ terms.
In terms of the coherent state representation discussed above, (\ref{renyi rho plus}) tells us that $\Tr\rho_A^n=\Tr\rho_+^n$, where $\rho_+$ is an operator whose generic matrix element reads
\be
\label{rhoplus coherent}
\rho_+(\zeta_A,\eta_A)       \,= \, \frac{1}{2}\Big[\rho_1(\zeta_A,\eta_A)  + \rho_2(\zeta_A,\eta_A)  + \rho_3(\zeta_A,\eta_A)   -  \rho_4(\zeta_A,\eta_A)   \Big] \,,
\ee
being $\rho_j(\zeta_A,\eta_A) $ with $1\leqslant j\leqslant 4$ the matrix elements discussed in \S\ref{sec:rdm}.
The expression (\ref{rhoplus coherent}) is meaningful also in the continuum limit and therefore also in this regime the $n$-th moment of the reduced density matrix reads
 $\Tr\rho_A^n = 2^{-n} \sum_{\boldsymbol{q}} (-1)^{\#4}\, \Tr\big[\prod_{k=1}^n \rho_{q_k} \big] $, which is a sum containing $4^n$ terms.
Each of these terms is characterised by a vector $\boldsymbol q$ made by $n$ integers $q_i\in\{1,2,3,4\}$ and has the following form 
\be
\label{terms Ising Dirac}
\Tr\bigg[\,\prod_{k=1}^n \rho_{q_k} \bigg] 
\,=\, 
\int\prod_{k=1}^n D\conj\zeta_k \, D\zeta_k \;\rho_{q_1}(-\zeta_n,\zeta_1)\prod_{k=2}^n \rho_{q_k} (\zeta_{k-1},\zeta_k)
\,,
\ee
where in the r.h.s.\ the first expression in (\ref{eq:identity and trace}) has been employed $n-1$ times, while the trace according to the second expression in (\ref{eq:identity and trace}) has been taken only once. 
The expression (\ref{terms Ising Dirac}) provides the scaling limit of the lattice term in (\ref{renyi T_n}) characterised by the same $\boldsymbol{q}$.

At this point, it is straightforward to adapt to the continuum case the same symmetry considerations done to get \eqref{renyi T_n} on the lattice. 
In particular, two terms characterised by $\boldsymbol q$ and $\boldsymbol q'$ are equal if they are related by the exchange $1\leftrightarrow 2$ and $3\leftrightarrow 4$.
Furthermore, because of the relative minus sign in front of $\rho_3$ and $\rho_4$ in (\ref{rhoplus coherent}), terms with an odd total number of $\rho_3$'s and $\rho_4$'s cancel out in the sum for $\Tr\rho_+^n$, while terms with an even total number of $\rho_3$'s and $\rho_4$'s survive and they are pairwise equal.
The final result for $\Tr\rho_A^n$ has the same structure of \eqref{renyi T_n} and its terms contain the operator $(-1)^F$ along non trivial closed curves on $\mathcal{R}_n$.
These closed curves are around $A_2$ whenever $\rho_2$ or $\rho_4$ occur in (\ref{terms Ising Dirac})  and along $B_1$ on two different sheets if (\ref{terms Ising Dirac}) contains a couple of $\rho_3$'s, or a couple of $\rho_4$'s or the mixed combination (the latter closed curves are easier to visualise on the representation of the multisheet Riemann surface given e.g. in Fig.\,\ref{fig:cage} for the case $n=4$).
The $2^{2(n-1)}$ terms in $\Tr\rho_A^n$ correspond to all the inequivalent insertions of the operator $(-1)^F$ around the cut in $A_2$ or along two $B_1$'s on different sheets.
Indeed $\Tr\rho_A^n$ is given as a sum over all possible characteristics, or, equivalently, over all possible boundary conditions (either periodic or antiperiodic) along the homology cycles.

Since the occurrence of $(-1)^F$ on $\mathcal{R}_n$ influences the fermionic boundary conditions along the basis cycles, it is natural to look for a relation between the generic term  (\ref{terms Ising Dirac}) characterised by $\boldsymbol q$ and the partition function of the corresponding fermionic model on the Riemann surface $\mathcal R_n$ with a given spin structure, namely with the proper set of boundary conditions imposed along the $a$ and $b$ cycles.

As for the modular invariant Dirac fermion,
the contribution characterised by a fixed set of boundary conditions reads
\begin{equation} 
\label{corr: renyi xx}
 \Tr\bigg[ \, \prod_{k=1}^{n} \rho_{q_k} \bigg] 
 =
  \frac{\big(c_n^\text{XX}\big)^2}{\big[\ell_1\ell_2 (1-x)\big]^{2\Delta_n}} \, 
  \bigg|\frac{\Theta[ \boldsymbol e ] ({\bf 0}|\tau(x))}{\Theta({\bf 0}|\tau(x))}\bigg|^2 ,
\end{equation}
where $c=1$, $\Delta_n$ is given by (\ref{Delta_n}) and  $c_n^\text{XX}$ by (\ref{c_n free fermion});
while for the Ising model we have
\begin{equation} 
\label{corr: renyi ising}
 \Tr\bigg[  \,\prod_{k=1}^{n} \rho_{q_k} \bigg] 
 =
    \frac{\big(c_n^\text{Ising}\big)^2}{\big[\ell_1\ell_2 (1-x)\big]^{2\Delta_n}} \, 
  \bigg|\frac{\Theta[ \boldsymbol e ] 
({\bf 0}|\tau(x))}{\Theta({\bf 0}|\tau(x))}\bigg| \,,
\end{equation}
where $c=1/2$ and $c_n^\text{Ising}$ is given by  (\ref{c_n ising}).

In order to complete the identifications (\ref{corr: renyi xx}) and (\ref{corr: renyi ising}), in the following we provide a rule to associate the  spin structure $\boldsymbol e$ in the r.h.s.'s of (\ref{corr: renyi xx}) and (\ref{corr: renyi ising}) to the vector $\boldsymbol q$ in the l.h.s.'s.
Such rule depends on the canonical homology basis chosen to write the period matrix of $\mathcal R_n$ and in our case it is given by the cycles shown in Figs.\,\ref{fig:cage} and \ref{fig:R4} for the case $n=4$ \cite{ctt-14}.
A vector $\boldsymbol q$ allows us to find the closed curves where the operator $(-1)^F$ must be inserted on $\mathcal R_n$ as explained above.
The operators $(-1)^F$ occurring in our analysis can be distinguished in two types: $(-1)^{F_{A_2}}$  and $(-1)^{F_{B_1}}$, depending on whether they come from $P_{A_2}$ or $P_{B_1}$.
Considering the cycles $\{a_r, b_r \,; \, 1\leqslant r\leqslant n-1\}$ of the canonical homology basis, we have to count how many times they cross the curves where $(-1)^F$ occurs.
The parity of these numbers provide the characteristic corresponding to $\boldsymbol q$ as follows:
if $a_k$ crosses the curves where $(-1)^F$ is inserted an even number of times, then $\varepsilon_k=0$ (antiperiodic b.c.\ along $a_k$), otherwise $\varepsilon_k=1/2$ (periodic b.c.\ along $a_k$).
Notice that the $a$ cycles can meet only operators $(-1)^{F_{B_1}}$, i.e. the ones along the $B_1$'s of the various sheets.
Similarly, if $b_k$ crosses the curves along which $(-1)^F$ is placed an even number of times, we have $\delta_k=0$ (antiperiodic b.c.\ along $b_k$), otherwise $\delta_k=1/2$ (periodic b.c.\ along $b_k$).
We remark that the $b$ cycles cross only operators $(-1)^{F_{A_2}}$, namely the ones placed along the edges of the slits in $A_2$.

The map between $\boldsymbol q$ and the corresponding characteristic $\boldsymbol e$ just discussed can be written more explicitly. 
Considering the  $j$-th sheet ($1\leqslant j \leqslant n$), let us introduce
\begin{equation}
\label{p_j 01}
p^{B_1}_j =
\begin{cases}
0 & \text{if $(-1)^{F_{B_1}}$ does not occur} \\
1 & \text{if $(-1)^{F_{B_1}}$ occurs,}
\end{cases}
\qquad
p^{A_2}_j =
\begin{cases}
0 & \text{if $(-1)^{F_{A_2}}$ does not occur} \\
1 & \text{if $(-1)^{F_{A_2}}$ occurs,}
\end{cases}
\end{equation}
which  can be expressed also in a closed form as $p^{B_1}_j = \lfloor(q_j-1)/2\rfloor$ and $p^{A_2}_j = (1+(-1)^{q_j})/2$,
where $q_j$ is the $j$-th element of $\boldsymbol q$ and we denoted by $\lfloor x \rfloor$ the integer part of $x$.
Then, the characteristic $\boldsymbol e$ associated to $\boldsymbol q$ in \eqref{corr: renyi xx} or \eqref{corr: renyi ising}  is given by
\be
\label{corr renyi eps}
\begin{array}{l}
\displaystyle 
2\,\varepsilon_k = \bigg(\sum_{\ell=1}^{k}p_\ell^{B_1}\bigg)\textrm{ mod $2$} \,=\, \big[1-(-1)^{\sum_{\ell=1}^{k} p_\ell^{B_1}}\big]/2 \,, 
\\
\rule{0pt}{.7cm}
\displaystyle 
2\, \delta_k = \left(p^{A_2}_k + p_{k+1}^{A_2}\right) \textrm{ mod $2$} \,=\, \big[1-(-1)^{p^{A_2}_k + p_{k+1}^{A_2}}\big]/2 \,,
\end{array}
\ee
where $1\leqslant k \leqslant n-1$.
This relation between the vector $\boldsymbol q$ and the spin structure $\boldsymbol e$ shows that each term \eqref{corr: renyi xx} or \eqref{corr: renyi ising} is the scaling limit of the term characterised by the same $\boldsymbol q$ in the expression \eqref{renyi T_n} for $\Tr\rho_A^n$ for the lattice.

Since a Riemann theta function with odd characteristic vanishes identically in the expressions (\ref{Fn Dirac}) and (\ref{Fn Ising}), an interesting consequence of this analysis 
is that the scaling limit of the terms in \eqref{renyi T_n} associated to odd characteristics through the rules discussed above is zero identically.
These terms turn out to be the ones with an odd number of $\rho_4$'s, namely the ones occurring in \eqref{renyi T_n} with a minus sign.
This observation about the odd spin structures is independent of the choice of the homology basis on $\mathcal R_n$.
Indeed, two canonical homology basis are related by a modular transformation and this kind of maps leaves invariant the parity of the characteristics.

\subsection{Partial transpose of the reduced density matrix}
\label{sec:pt2 coherent states}

\begin{figure}
\vspace{.6cm}
\includegraphics[width=1.\textwidth]{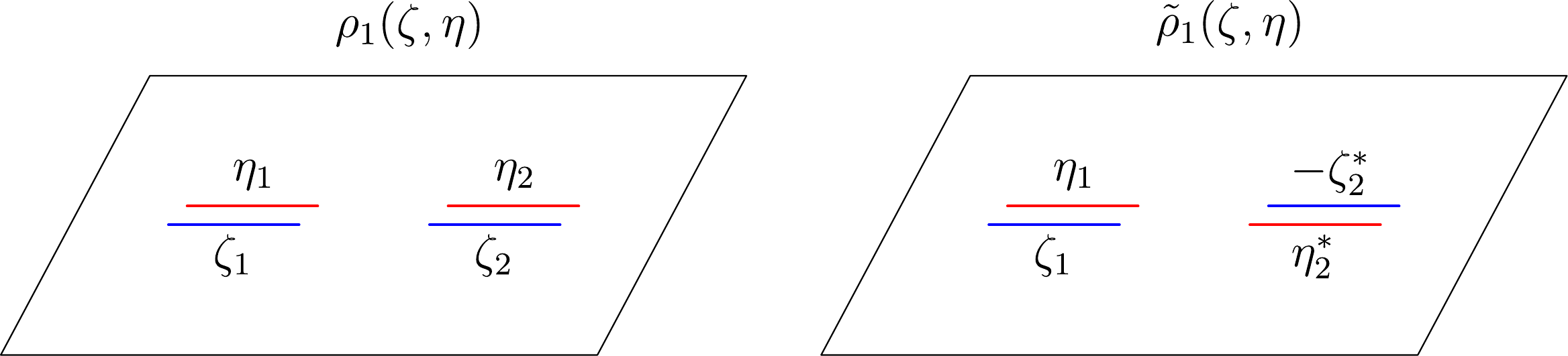}
\vspace{-.3cm}
\caption{
Path integral representation in the coherent state basis of $\rho_1(\zeta,\eta)$ (left panel and top left panel of Fig.\,\ref{fig:shee}) occurring in $\Tr\rho_A^n$ and of $\tilde{\rho}_1(\zeta,\eta)$ 
occurring in $\Tr(\rho^{T_2}_A)^n$.
The joining conditions along $B$ are the same ones shown top left panel of Fig.\,\ref{fig:shee}.
The path integral representations of $\tilde{\rho}_j(\zeta,\eta)$ for $j\in \{2,3,4\}$ can be depicted starting from the remaining panels of Fig.\,\ref{fig:shee} and performing the same change in the boundary conditions along the edges of $A_2$ shown here. 
}
\label{fig:sheetstilde}
\end{figure}

In this subsection we represent the matrix elements of the matrices (\ref{rho tilde 1234 def}) through the coherent states, as done in \S\ref{sec:rdm} for the matrices (\ref{rho 1234 def}). 
Again, their generalisation to the continuum regime is straightforward.

In \cite{ctc-15-2}  the path integral representation of $\tilde{\rho}_1$ and $\tilde{\rho}_2$ in \eqref{rhoA T2 1234} has been discussed in detail, finding that  the coherent states matrix element $\tilde \rho_1(\zeta_A,\eta_A)$ reads
\bea
\label{eq:rhotilde1 path integral}
& &
\hspace{-2.2cm}
\tilde \rho_1(\zeta_A,\eta_A) \,= \, 
  e^{-\conj\zeta_A \eta_A}   \bra{\zeta_{A_1}, \eta^\ast_{A_2} }V_2 \, \rho_1   V_2^\dagger \ket{\eta_{A_1} ,  -\zeta^\ast_{A_2}}
 \\
 & & \hspace{-2.2cm}
  =\,  e^{-\conj\zeta_A \eta_A} 
 \int D\chi_B^\ast \, D\chi_B
 \, e^{-\conj\chi_B \,\chi_B} \braket{\zeta_{A_1}, \eta^\ast_{A_2}; -\chi_{B_1}, -\chi_{B_2}  | V_2 |\Psi}  \braket{\Psi| V_2^\dagger| \eta_{A_1} ,  -\zeta^\ast_{A_2};\,\chi_{B_1},\chi_{B_2}},
 \nonumber
\eea
where $V_2$ is the unitary operator $ V_2 \equiv U_{-\pi}^{(y)} \, U_{-\pi/2}^{(z)}$ (see (\ref{Uk def}) for the definition of $U^{(k)}_\alpha$), which  sends $a^x_j\rightarrow -a^y_j$ and $a^y_j\rightarrow -a^x_j$, for $j\in A_2$.
In order to compute the matrix element $\tilde \rho_2(\zeta_A,\eta_A)$, the effect of $P_{A_2}$ discussed in \S\ref{sec:rdm} must be taken into account first (we recall that $\iu a^xa^y\ket{\zeta}=-\ket{\zeta}$) and then \eqref{eq:rhotilde1 path integral} can be employed. 
The result is
\bea
\label{eq:rhotilde2 path integral}
& &  \hspace{-2.5cm}
 \tilde{\rho}_2(\zeta_A,\eta_A) \,= \, 
  e^{-\conj\zeta_A \eta_A}   \bra{\zeta_{A_1}, \zeta_{A_2} } P_{A_2} \tilde{\rho}_1   P_{A_2}   \ket{\eta_{A_1} ,  \eta_{A_2}}
  \,=\,
  e^{-\conj\zeta_A \eta_A}   \bra{\zeta_{A_1}, -\zeta_{A_2} }  \tilde{\rho}_1    \ket{\eta_{A_1} ,  -\eta_{A_2}}
 \\
& & \hspace{-2.5cm}
  =\,  e^{-\conj\zeta_A \eta_A} 
 \int D\chi_B^\ast \, D\chi_B
 \, e^{-\conj\chi_B \,\chi_B} \braket{\zeta_{A_1}, -\eta^\ast_{A_2}; -\chi_{B_1}, -\chi_{B_2}  | V_2 |\Psi}  \braket{\Psi| V_2^\dagger| \eta_{A_1} ,  \zeta^\ast_{A_2};\,\chi_{B_1},\chi_{B_2}} .
 \nonumber
\eea
The first step in (\ref{eq:rhotilde2 path integral}) is just the definition, in the second one the effect of $P_{A_2}$ has been implemented and in the last one \eqref{eq:rhotilde1 path integral} has been applied. 
Equivalently, we can observe that $ \tilde{\rho}_2(\zeta_A,\eta_A) =   \tilde{\rho}_1(\zeta_{A_1},-\zeta_{A_2};\eta_{A_1},-\eta_{A_2}) $.

As for the matrix element $\tilde\rho_3(\zeta_A,\eta_A)$, from \eqref{rhoA tilde B1 def} it is clear that the difference between $\tilde\rho_1$ and $\tilde\rho_3$ is the occurrence of the string $P_{B_1}$ in the coefficients $w_{12}^{B_1}$ (see (\ref{rho even def}) and (\ref{rho odd def})).
Since all the steps done in \cite{ctc-15-2} to compute $\tilde\rho_1(\zeta_A,\eta_A)$ involve only the fermions in $A_2$ and they are independent of the values of the coefficients $w_{12}$, we can perform the same kind of computations for $\tilde\rho_3(\zeta_A,\eta_A)$, finding that
\bea
\label{eq:rhotilde3 path integral}
& &  \hspace{-2.3cm}
 \tilde{\rho}_3(\zeta_A,\eta_A) \,= \, 
  e^{-\conj\zeta_A \eta_A}   \bra{\zeta_{A_1}, \eta^\ast_{A_2} }V_2  \rho_3     V_2^\dagger \ket{\eta_{A_1} ,  -\zeta^\ast_{A_2}}
 \\
& & \hspace{-2.3cm}
=\, 
 e^{-\conj\zeta_A \eta_A} 
 \int D\chi_B^\ast \, D\chi_B
 \, e^{-\conj\chi_B \,\chi_B} \braket{\zeta_{A_1}, \eta^\ast_{A_2}; \chi_{B_1}, -\chi_{B_2}| V_2 |\Psi}  \braket{\Psi| V_2^\dagger |\eta_{A_1} , - \zeta^\ast_{A_2};\,\chi_{B_1},\chi_{B_2}}.
 \nonumber
\eea
The matrix element $\tilde\rho_4(\zeta_A,\eta_A)$ can be computed like in \eqref{eq:rhotilde3 path integral}, by taking into account also the occurrence of the string $P_{A_2}$, as done in getting (\ref{eq:rhotilde2 path integral}) from (\ref{eq:rhotilde1 path integral}). The result reads
\bea
\label{eq:rhotilde4 path integral}
& &  \hspace{-2.5cm}
 \tilde{\rho}_4(\zeta_A,\eta_A) \,= \,  
 e^{-\conj\zeta_A \eta_A}   \bra{\zeta_{A_1}, \zeta_{A_2} } P_{A_2} \tilde{\rho}_3   P_{A_2}   \ket{\eta_{A_1} ,  \eta_{A_2}}
  \,=\,
  e^{-\conj\zeta_A \eta_A}   \bra{\zeta_{A_1}, -\zeta_{A_2} }  \tilde{\rho}_3    \ket{\eta_{A_1} ,  -\eta_{A_2}}
 \\
& & \hspace{-2.5cm}
=\, 
 e^{-\conj\zeta_A \eta_A} 
 \int D\chi_B^\ast \, D\chi_B
 \, e^{-\conj\chi_B \,\chi_B} \braket{\zeta_{A_1}, - \eta^\ast_{A_2}; \chi_{B_1}, -\chi_{B_2}| V_2 |\Psi}  \braket{\Psi| V_2^\dagger |\eta_{A_1} , \zeta^\ast_{A_2};\,\chi_{B_1},\chi_{B_2}},
 \nonumber
\eea
which can be obtained also from (\ref{eq:rhotilde3 path integral}), since $\tilde\rho_4(\zeta_A,\eta_A) = \tilde\rho_3(\zeta_{A_1},-\zeta_{A_2};\eta_{A_1},-\eta_{A_2})$.

The continuum version of \eqref{eq:rhotilde1 path integral}, \eqref{eq:rhotilde2 path integral}, \eqref{eq:rhotilde3 path integral} and \eqref{eq:rhotilde4 path integral} can be found exactly as discussed in \S\ref{sec:rdm} to get continuum version of  \eqref{eq:rho1 path integral}, \eqref{eq:rho2 path integral}, \eqref{eq:rho3 path integral} and \eqref{eq:rho4 path integral}.
Besides the occurrence of the operator $V_2$, the main difference between $ \rho_j(\zeta_A,\eta_A)$ and $ \tilde{\rho}_j(\zeta_A,\eta_A)$ is given by the fact that the fermionic fields above and below the cut in $A_2$ must be exchanged. 
This is represented pictorially in Fig.\,\ref{fig:sheetstilde} for $\tilde\rho_1(\zeta_A,\eta_A)$.

The presence of $V_2$ turns out to be irrelevant.
Indeed, in~\cite{ctc-15-2} it was found that it can be written in terms of the fermionic operators as
$V_2=\exp\big\{\sum_{j\in A_2}[-\iu\frac{\pi}{2} (e^{\iu\frac{\pi}{4}}c_j^\dag+e^{-\iu\frac{\pi}{4}}c_j )]\big\}$.
When taking the continuum limit of \eqref{eq:rhotilde1 path integral}, \eqref{eq:rhotilde2 path integral}, \eqref{eq:rhotilde3 path integral} and \eqref{eq:rhotilde4 path integral}, 
the presence of $V_2$ translates into a contribution to the action which is linear in the fermionic field and localized both above and below $A_2$.
Such term can be eliminated by a redefinition of the field.
This fact reflects the freedom in defining the partial transposition of a fermionic operator on the lattice \eqref{hat transposition def} modulo some unitary transformation \cite{ez-15}.
Indeed, an equivalent way to define the partial transposition could have been easily introduced in order to cancel the contribution of $V_2$.

\subsection{Moments of the partial transpose}
\label{sec:moments pt2}

In this subsection we adapt the analysis performed in \S\ref{sec:renyi} for the moments $\Tr \rho_A^n$ to the moments $\Tr (\rho_A^{T_2})^n$,
finding that the scaling limit of the term associated to the vector $\tilde{\boldsymbol{q}}$ in the sum (\ref{ptA2 tilde T_n})  is the term characterised by a specific spin structure $\boldsymbol e$  in the sums  (\ref{tilde Fn Dirac})  and (\ref{tilde Fn Ising}).

In \S\ref{subsec pt2 lattice} we have seen that $\Tr (\rho_A^{T_2})^n=\Tr(\rho_+^{T_2})^n$ (see (\ref{traces n rhopm T2})).
In terms of coherent states, the generic matrix element of the operator $\rho_+^{T_2}$ reads
\be
\rho_+^{T_2}(\zeta,\eta) \,=\, \frac{1}{2}\Big[\tilde\rho_1(\zeta,\eta) + \tilde\rho_2(\zeta,\eta) -\iu\, \tilde\rho_3(\zeta,\eta) +\iu\, \tilde\rho_4(\zeta,\eta)\Big] \,,
\ee
where $\tilde\rho_j(\zeta_A,\eta_A)$ have been written explicitly in \S\ref{sec:pt2 coherent states}.
Thus, the $n$-th moment becomes $\Tr (\rho_A^{T_2})^n= 2^{-n} \sum_{\tilde{\boldsymbol{q}}} (-\iu)^{\#3} \, \iu^{\#4} \, \Tr\big[\prod_{k=1}^n \tilde\rho_{\tilde{q}_k} \big] $, where the sum contains $4^n$ terms and each of them is identified by the $n$ dimensional vector $\tilde{\boldsymbol{q}}$ whose elements are $\tilde{q}_i \in \{1,2,3,4\}$.
The term characterised by the vector $\tilde{\boldsymbol{q}}$  in the above sum providing the moments of the partial transpose can be written as follows
\be 
\label{PT2 generic term}
\Tr\bigg[\,\prod_{k=1}^n \tilde\rho_{\tilde q_k} \bigg] 
\,=\, 
\int\prod_{k=1}^n D\conj\zeta_kD\zeta_k \,\tilde\rho_{\tilde q_1}(-\zeta_n,\zeta_1)\prod_{k=2}^n \tilde \rho_{\tilde q_k} (\zeta_{k-1},\zeta_k)\,,
\ee
which resembles the generic term of $\Tr \rho_A^n$ given in (\ref{terms Ising Dirac}).
As discussed after \eqref{terms Ising Dirac}, also in this case it is immediate to adapt the symmetry considerations discussed for the corresponding lattice formula \eqref{ptA2 tilde T_n}. 
Again, two terms characterised by $\tilde{\boldsymbol q}$ and $\tilde{\boldsymbol q}'$ are equal if they are related by the exchange $1\leftrightarrow 2$ and $3\leftrightarrow 4$.
The same cancellations of the terms with an odd total number of $\rho_3$'s and $\rho_4$'s happen, and the remaining terms are the ones where the operator $(-1)^F$ is inserted along non trivial closed curves on $\widetilde{\mathcal{R}}_n$, either above and below the cut in $A_2$ or along $B_1$ on two different sheets.
The final result for $\Tr (\rho_A^{T_2})^n$ has  the same form of \eqref{ptA2 tilde T_n}, and the sum is performed over all the inequivalent insertions of the operator $(-1)^F$ on the different sheets.
This is equivalent to a sum over all possible boundary conditions along the basis cycles  of $\widetilde{\mathcal{R}}_n$.

Mimicking the analysis performed in \S\ref{sec:renyi} for $\Tr \rho_A^n $, we want to interpret the generic term (\ref{PT2 generic term}) as the partition function of the fermionic model on the Riemann 
surface $\widetilde{\mathcal R}_n$ with the proper spin structure, that can be determined uniquely from the vector $\tilde{\boldsymbol{q}}$.
For the CFT models we are considering, the moments of the partial transpose $\Tr (\rho_A^{T_2})^n$ have been written in  (\ref{tr rhoAn T2 cft}) and (\ref{G via F}), with the functions $\mathcal{F}_n(\tfrac{x}{x-1})$ given by (\ref{tilde Fn Dirac}) and (\ref{tilde Fn Ising}) as sums over all possible spin structures. 
Thus, for the modular invariant Dirac fermion, we have
\begin{equation} 
\label{corr: neg xx}
 \Tr\bigg[ \, \prod_{k=1}^{n} \tilde\rho_{\tilde q_k} \bigg] 
 =
  \big(c_n^\text{XX}\big)^2
  \left(\frac{1-x}{\ell_1\ell_2 }\right)^{2\Delta_n} \, \bigg|\frac{\Theta[ \boldsymbol e ] 
({\bf 0}|\tilde\tau(x))}{\Theta({\bf 0}|\tilde\tau(x))}\bigg|^2 ,
\end{equation}
where $\Delta_n$ is given by (\ref{Delta_n}) with $c=1$ and the non universal constant $c_n^\text{XX}$ is \eqref{c_n free fermion}.
Similarly, for the Ising model, we find
\begin{equation} 
\label{corr: neg ising}
 \Tr\bigg[  \,\prod_{k=1}^{n} \tilde\rho_{\tilde q_k} \bigg] 
 =
  \big(c_n^\text{Ising}\big)^2
  \left(\frac{1-x}{\ell_1\ell_2 }\right)^{2\Delta_n} \, \bigg|\frac{\Theta[ \boldsymbol e ] 
({\bf 0}|\tilde\tau(x))}{\Theta({\bf 0}|\tilde\tau(x))}\bigg| \,,
\end{equation}
where $c=1/2$ and $c_n^\text{Ising}$ is given by \eqref{c_n ising}.
In order to complete the identifications (\ref{corr: neg xx}) and (\ref{corr: neg ising}), in the following we provide a bijective relation between the vector $\tilde{\boldsymbol q}$ in the l.h.s.'s and the characteristic $\boldsymbol e$ in the r.h.s.'s, namely a map between the terms in the sum \eqref{ptA2 tilde T_n} and the set of the possible spin structures.

The relation between $\tilde{\boldsymbol q}$ and the characteristic $\boldsymbol e$ is similar to the one discussed for $\Tr\rho_A^n$ in \S\ref{sec:renyi}.
The vector $\tilde{\boldsymbol q}$ contains all the information to place the operator $(-1)^F$ along some closed curves on $\widetilde{\mathcal R}_n$.
Then, considering the canonical homology basis $\{\tilde{a}_r, \tilde{b}_r \,; \, 1\leqslant r\leqslant n-1\}$, in order to find the characteristic associated to $\tilde{\boldsymbol q}$ we have to count how many times each cycle of the basis intersects the closed curves supporting the operator $(-1)^F$. 
The parity of this number for the cycle $\tilde{a}_k$ provides $\varepsilon_k$ (where $1\leqslant k \leqslant n-1$) as discussed in \S\ref{sec:renyi} (an even number of times corresponds to $\varepsilon_k=0$, i.e. to antiperiodic b.c.\ along $a_k$, while $\varepsilon_k=1/2$, meaning periodic b.c.\ along $a_k$, when this number is odd), while the parity of the number associated to the cycle $\tilde{b}_r$ determines $\delta_r$ in a similar way.

This relation between $\tilde{\boldsymbol q}$ and  $\boldsymbol e$ can be written more explicitly. 
By introducing $\tilde p^{B_1}_j$ and $\tilde p^{A_2}_j$ in terms of $\tilde q_j$ exactly as shown in (\ref{p_j 01}) for the corresponding untilded quantities,
the characteristic $\boldsymbol e$ associated to $\tilde{\boldsymbol q}$ in (\ref{corr: neg xx}) and (\ref{corr: neg ising}) reads
\be
\label{corr pt2 moments}
%\left\{ \rule{0pt}{1cm} \right.
\begin{array}{l}
\displaystyle 
2\,\varepsilon_k = \bigg(\sum_{\ell=1}^{k}\tilde p_\ell^{B_1}\bigg)\textrm{ mod $2$} \,=\, \big[1-(-1)^{\sum_{\ell=1}^{k}\tilde p_\ell^{B_1}}\big]/2 \,, 
\\
\rule{0pt}{.7cm}
\displaystyle 
2\,\delta_k= \big(\tilde p^{B_1}_j + \tilde p^{A_2}_j + \tilde p_{j+1}^{A_2}\big)\textrm{ mod $2$}
\,=\, 
\big[1-(-1)^{\tilde p^{B_1}_k + \tilde p^{A_2}_k + \tilde p_{k+1}^{A_2}}\big]/2 \,.
\end{array}
\ee
Also in the case of $\Tr(\rho_A^{T_2})^n$ we find that the terms on the lattice occurring in the sum \eqref{ptA2 tilde T_n} with a minus sign correspond via \eqref{corr: neg xx} or \eqref{corr: neg ising} to terms whose Riemann theta functions have odd characteristics; hence they vanish in the scaling limit.

Since the two Riemann surfaces $\mathcal{R}_n$ and $\widetilde{\mathcal R}_n$ are different, also the ways to associate a characteristic $\boldsymbol e$ to the vectors  $\boldsymbol q$ and $\tilde{\boldsymbol q}$ are different, despite the fact that the underlying principle is the same. 
An important distinction  to remark between $\Tr\rho_A^n$ and $\Tr(\rho_A^{T_2})^n$ is that in the latter case the cycle $\tilde b_k$ crosses $B_1$ on the $k$-th sheet (see Fig.\,\ref{fig:bcycles}, right panel).
Thus, according to the rule relating $\tilde{\boldsymbol q}$ and $\boldsymbol e$ explained above, we have that $\boldsymbol\delta$ is determined also by the occurrence of $(-1)^F$ along $B_1$, and not only around the cut in $A_2$, like in the case of $\Tr\rho_A^n$ (compare the formulas for $\delta_j$ in (\ref{corr renyi eps}) and (\ref{corr pt2 moments})).

An analysis similar to the one presented here has been carried out in \cite{ctc-15-2} for the free fermion. 
The main difference between the free fermion and the critical XX model on the lattice is the occurrence of the string of Majorana operators $P_{B_1}$ along the sites of $B_1$ separating the two blocks $A_1$ and $A_2$, which corresponds to the operator $(-1)^{F_{B_1}}$ in the scaling limit. 
This implies that for the modular invariant Dirac fermion the boundary conditions along the $a$ cycles of $\widetilde{\mathcal{R}}_n$ may be periodic or antiperiodic, while for the free fermion the boundary conditions along the $a$ cycles are all antiperiodic.

By eliminating the operator $P_{B_1}$ (or $(-1)^{F_{B_1}}$  in the scaling regime) from the results for the critical XX model one gets the corresponding ones for the free fermion. 
This can be done both on the expressions on the lattice and on the ones in the scaling limit. 
In the appendix \S\ref{sec:free fermion} we briefly illustrate this procedure for both $ \Tr \rho_A^n $ and $ \Tr ( \rho_A^{T_2})^n $.

\section{Numerical checks}
\label{sec: numerics xy}

The CFT predictions for $ \Tr\rho_A^n$ and $ \Tr ( \rho_A^{T_2})^n$  given in \eqref{Fn Dirac}, \eqref{Fn Ising}, \eqref{tilde Fn Dirac} and \eqref{tilde Fn Ising}
have been already checked for $n=3,4,5$  through lattice computations on the XX spin chain  and Ising chain at criticality \cite{fc-10,  ctt-14, a-13, ctc-15}.

In \cite{ctc-15-2} the moments of the partial transpose for the free Dirac fermion have been studied starting from the lattice formulation and by employing the corresponding coherent state path integral for the continuum limit.
In this case, the most important factor in the final formula for $ \Tr ( \rho_A^{T_2})^n$ is similar to the r.h.s.\ of (\ref{tilde Fn Dirac}): the generic term in the sum is the same but the sum is performed over the  characteristics with $\boldsymbol{\varepsilon} = \boldsymbol{0} $ and the coefficients of the various terms are different. 
The procedure developed for the free Dirac fermion in \cite{ctc-15-2} shows how each term in the scaling formula for $ \Tr ( \rho_A^{T_2})^n$ can be recovered as the continuum limit of its lattice counterpart.
In this paper we have extended this term-by-term correspondence to the case of $ \Tr\rho_A^n$ and $ \Tr ( \rho_A^{T_2})^n$ for the modular invariant Dirac fermion and the Ising model.
In this section we provide explicit numerical evidence of these term-by-term relations in the simplest cases of $n=2$ and $n=3$.

Each term in \eqref{renyi T_n} and \eqref{ptA2 tilde T_n} is the trace of the product of $n$ matrices taken among the ones in \eqref{rho 1234 def} and \eqref{rho tilde 1234 def} respectively.
In order to evaluate these terms, we employ the techniques developed in \cite{fc-10} for $\Tr\rho_A^n$ and then generalised in \cite{ctc-15} to compute $ \Tr ( \rho_A^{T_2})^n $.
In \cite{ctc-15-2} this analysis has been done for the free fermion. 
Since the Hamiltonians in \eqref{eq:HXX} and \eqref{eq:Hising} are quadratic in the fermionic operators, the ground state fermionic reduced density 
matrix $\rho_A^{\,\boldsymbol 1}=\rho_1$ in (\ref{rho 1234 def}) is Gaussian.
As for $\tilde\rho_A^{\,\boldsymbol 1}$, from \eqref{rhoA tilde B1 def} one can easily observe that it is also Gaussian. 
It is worth remarking that the operator $P_C$ (in our problem only $P_{A_2}$ and $P_{B_1}$ occur) does not spoil Gaussianity; indeed, it can be written as the exponential of a quadratic operator.
This implies that also the other matrices $\rho_k$ and $\tilde{\rho}_k$ for $k\in \{2,3,4\}$ in \eqref{rho 1234 def} and \eqref{rho tilde 1234 def} respectively are Gaussian.
Unfortunately, neither the four matrices $\rho_k$'s nor the $\tilde\rho_k$'s can be simultaneously diagonalized, and therefore the reduced density matrix \eqref{rhoA 1234} and its partial transpose \eqref{rhoA T2 1234} are not Gaussian.
This fact does not allow us to get the spectrum of $\rho_A$ and $\rho_A^{T_2}$, which would have provided the entanglement entropy and the logarithmic negativity respectively.
Nevertheless, in principle we can compute their moments for any value of $n$.
Similar considerations apply to the partial transpose for the free Dirac fermion as well \cite{ez-15}. 
On the contrary, for bosonic models considered in the literature, both $\rho_A$ and $\rho_A^{T_2}$ are Gaussian \cite{aepw-02,br-04,pe-09,mrpr-09} and this simplifies a lot the analysis.

In \eqref{eq:TrrhoA n2}, \eqref{eq:TrrhoA n3}, \eqref{eq:TrrhoAT n2} and \eqref{eq:TrrhoAT n3} the moments of $\rho_A$ and $\rho_A^{T_2}$ on the lattice have been written explicitly for $n=2$ and $n=3$.
Since for Gaussian states all the information of the system is encoded in the correlation matrices, 
the moments of $\rho_A$ and $\rho_A^{T_2}$  can be evaluated in a polynomial time in terms of the total size of the subsystem.
In particular, in our case the correlation matrices of the $\rho_k$'s and the $\tilde\rho_k$'s can be obtained from the ones corresponding to subsystems $A$ and $B_1$ as explained in \cite{fc-10, ctc-15}. 

The lattice computations have been performed in an infinite chain.
Equal size disjoint blocks $A_1$ and $A_2$ have been chosen ($\ell_1 = \ell_2 \equiv \ell$), separated by the block $B_1$ whose size is denoted by $d$.
The scaling regime, where the CFT predictions hold, is approached by taking configurations with increasing $\ell$, while  the ratio $\ell/d$ is kept fixed.
Thus, the four point ratio (\ref{cross ratio def}) becomes $x=[\ell/(\ell+d)]^2$ and configurations with the same value of $\ell/d$ correspond to the same $x$.

\begin{figure}[p!]
\hspace{-1cm}
\includegraphics[width=1.08\textwidth]{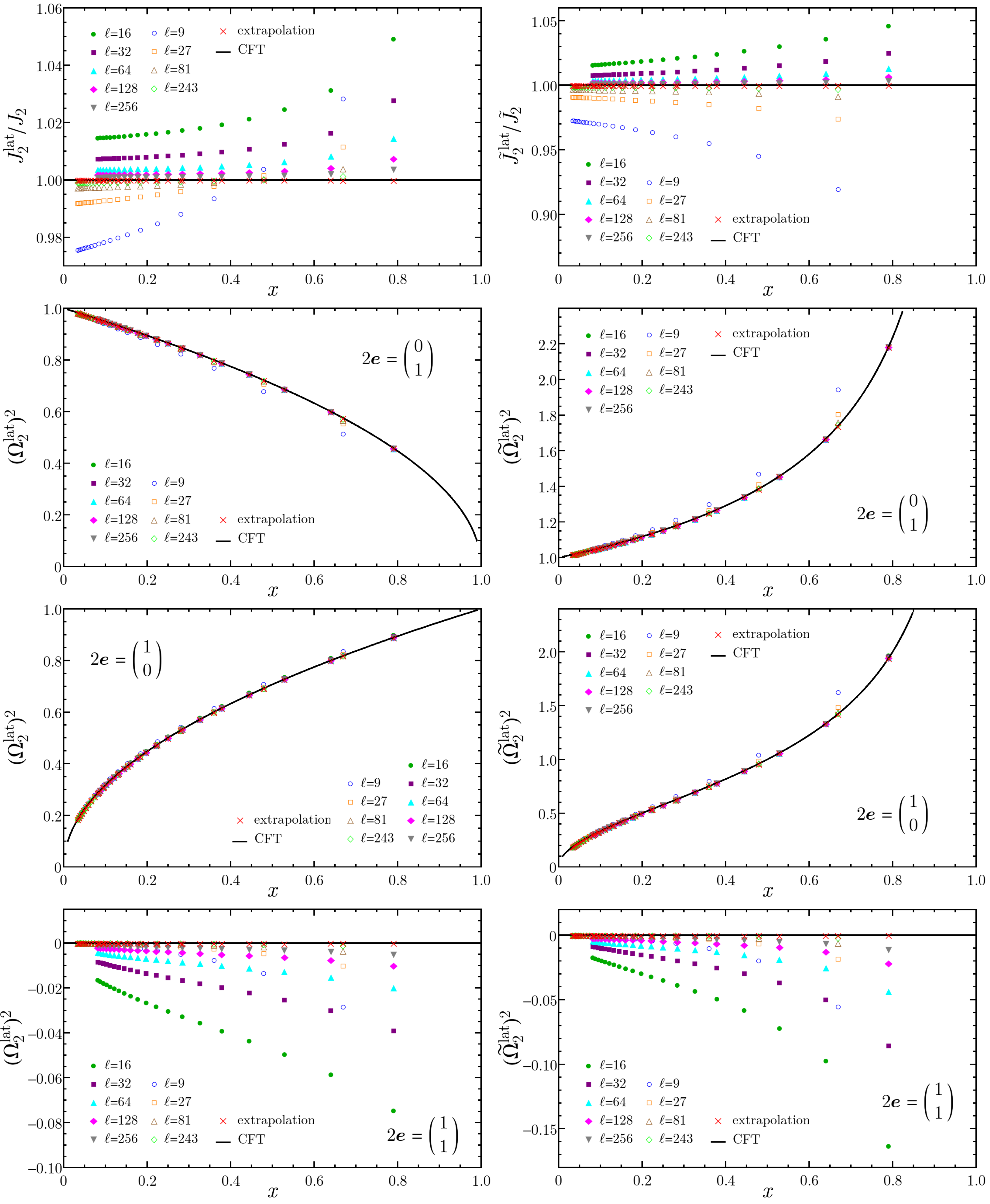}
\caption{
The terms occurring in $\Tr \rho_A^2$ (left panels) and $\Tr\big(\rho_A^{T_2}\big)^2$ (right panels) for the critical XX model (see (\ref{eq:TrrhoA n2}) and (\ref{eq:TrrhoAT n2}) respectively).
The extrapolated points (red crosses) are obtained through a fit of the data according to the scaling function \eqref{eq:scaling Omega XX} and they agree very well with the CFT predictions  (\ref{corr: renyi xx}) (solid lines).
} 
\label{fig:XX n2 all}
\end{figure}

\begin{figure}[p!]
\hspace{-0.cm}
\includegraphics[width=0.95\textwidth]{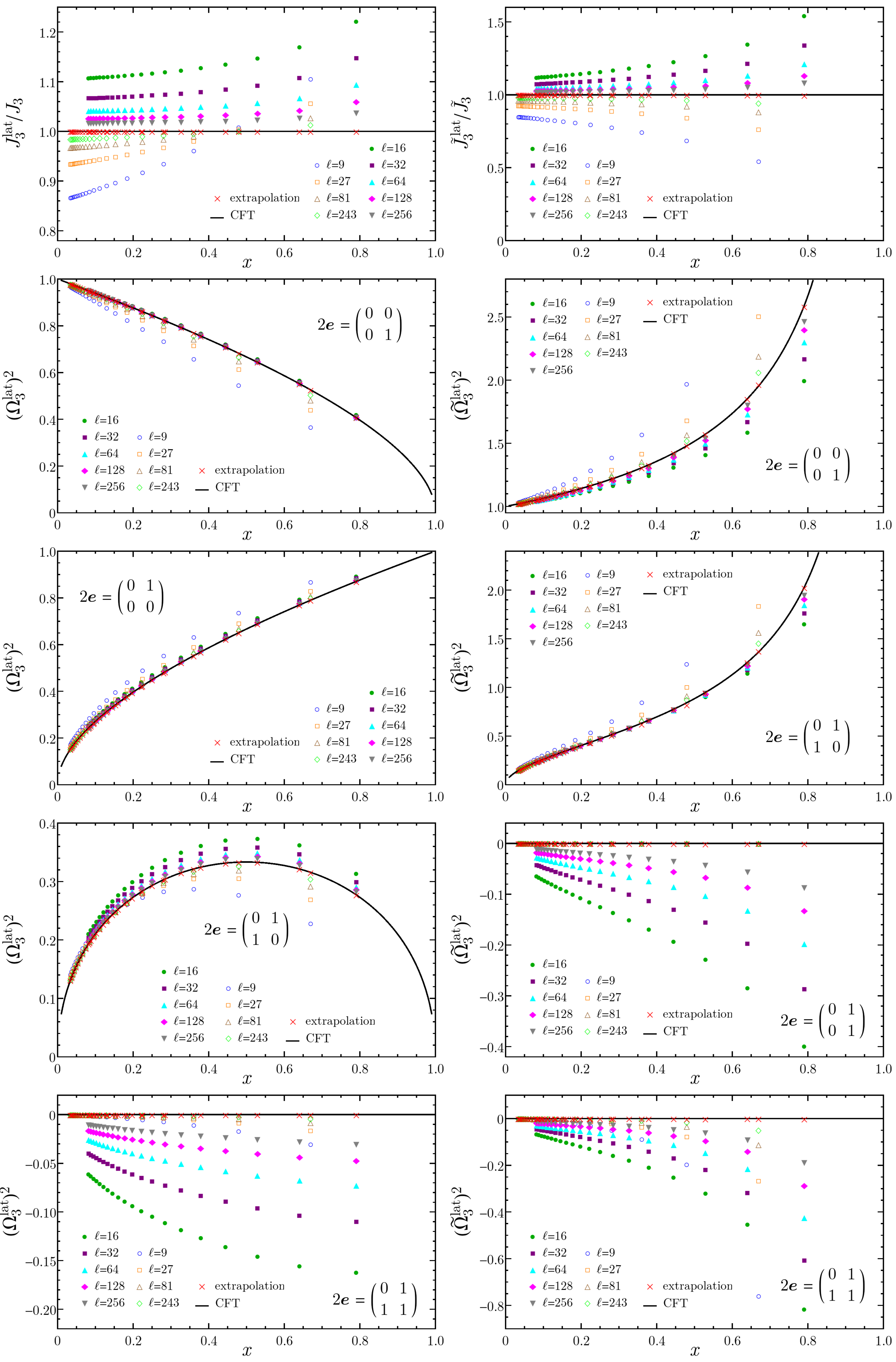}
\vspace{-.3cm}
\caption{
The terms in $\Tr \rho_A^3$ (left) and $\Tr\big(\rho_A^{T_2}\big)^3$ (right) for the critical XX model (see (\ref{eq:TrrhoA n3}) and (\ref{eq:TrrhoAT n3})  respectively).
For each group of identical terms, only one representative is shown.
Further details are given in the caption of Fig.\,\ref{fig:XX n2 all}.
} 
\label{fig:XX n3 all}
\end{figure}

Let us introduce the following lattice quantities associated to the moments of the reduced density matrix (\ref{renyi T_n}) for the XX spin chain 
\be
\label{J def lattice XX}
J_n^\text{lat} = 
  \Tr  \rho_{1}^n \,,
\hspace{2cm}
\Omega_n^\text{lat}[\boldsymbol q]^2  =  \frac{(-1)^{\#4}}{\Tr \rho_{1}^n}\;
 \Tr\bigg[ \prod_{k=1}^{n} \rho_{q_k} \bigg] \,,
\ee
and for the Ising spin chain
\be
 \label{J def lattice ising}
  J_n^\text{lat} = 
  \Tr  \rho_{1}^n\,,
\hspace{2cm}
 \Omega_n^\text{lat}[\boldsymbol q]  =  \frac{(-1)^{\#4}}{\Tr  \rho_{1}^n}\;
 \Tr\bigg[ \prod_{k=1}^{n} \rho_{q_k} \bigg] \,.
\ee
This notation has been introduced to make a direct comparison with the corresponding quantities \eqref{Jn Omega def} in the continuum. 
However, let us stress that, despite the notation, different correlators occur in \eqref{J def lattice XX} and \eqref{J def lattice ising}.
In particular, the r.h.s.\ of the second expression in \eqref{J def lattice XX} is not the square of the r.h.s.\ of the second equality in \eqref{J def lattice ising}.

Analogous quantities can be defined for the terms occurring in the formula  \eqref{ptA2 tilde T_n} for the moments of the partial transpose, namely
\bea
\label{J tilde def lattice XX}
& &
 \tilde J_n^\text{lat} = 
  \Tr  \tilde{\rho}_{1}^n \,,
\hspace{2cm}
 \widetilde\Omega_n^\text{lat}[\tilde{\boldsymbol q}]^2  =  \frac{(-1)^{\frac{\#4-\#3}{2}}}{\Tr  \tilde{\rho}_{1}^n}\;
 \Tr\bigg[ \prod_{k=1}^{n}\tilde{\rho}_{\tilde q_k} \bigg] \,,
 \\
  \rule{0pt}{.8cm}
\label{J tilde def lattice ising}
& &
 \tilde J_n^\text{lat} = 
  \Tr  \tilde{\rho}_{1}^n \,,
  \hspace{2cm}
 \widetilde\Omega_n^\text{lat}[\tilde{\boldsymbol q}]  =  \frac{(-1)^{\frac{\#4-\#3}{2}}}{\Tr  \tilde{\rho}_{1}^n}\;
 \Tr\bigg[ \prod_{k=1}^{n}\tilde{\rho}_{\tilde q_k} \bigg] \,,
\eea
for the XX spin chain and the Ising model respectively,
which are suitable to make contact with the continuum quantities in \eqref{tilde Jn Omega def}.

All these lattice quantities can be evaluated as explained in \cite{ctc-15}.
The main result of this paper is to show that the scaling limit of (\ref{J def lattice XX}), (\ref{J def lattice ising}), (\ref{J tilde def lattice XX}) and (\ref{J tilde def lattice ising}) gives the corresponding CFT expressions \eqref{Jn Omega def} and \eqref{tilde Jn Omega def}, where the non universal factor $c_n$ in $J_n$ and $\tilde{J}_n$ is \eqref{c_n free fermion} for the XX model and \eqref{c_n ising} for the Ising model.
The expressions of $J_n$ and $\tilde{J}_n$ depend on the model also through the value of the central charge $c$ in $\Delta_n$.

In the sums \eqref{renyi T_n} and \eqref{ptA2 tilde T_n}, many terms are equal because of the properties of the trace discussed in \S\ref{subsec rho_A} and \S\ref{subsec pt2 lattice}.
In the continuum, this degeneracy is due to the dihedral symmetry of the underlying Riemann surfaces (see \S\ref{sec:dihedral symmetry riemann surf}).
Despite the implementation of the dihedral symmetry allows to restrict our attention to one element for each equivalence class, the number of terms to deal with increases very fast with $n$.

\begin{figure}[p!]
\hspace{-1cm}
\includegraphics[width=1.08\textwidth]{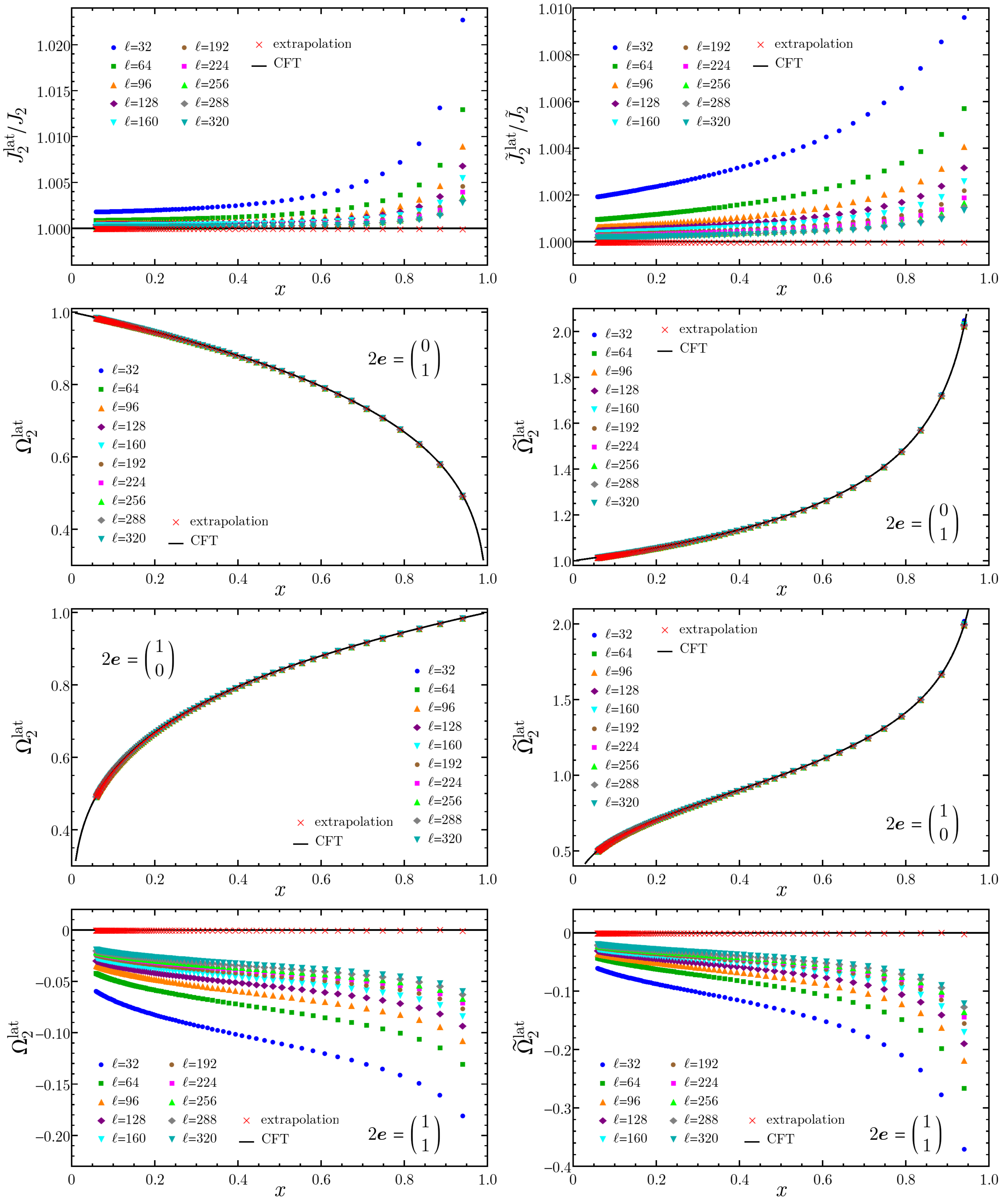}
\caption{
The terms occurring in $\Tr \rho_A^2$ (left panels) and $\Tr\big(\rho_A^{T_2}\big)^2$ (right panels) for the critical Ising chain (see (\ref{eq:TrrhoA n2}) and (\ref{eq:TrrhoAT n2}) respectively).
The extrapolated points (red crosses) are obtained through a fit of the data according to the scaling function \eqref{eq:scaling Omega Ising} and they agree very well with the CFT predictions  (\ref{corr: renyi ising}) (solid lines).
} 
\label{fig:Is n2 all}
\end{figure}

\begin{figure}[p!]
\hspace{-0.cm}
\includegraphics[width=0.95\textwidth]{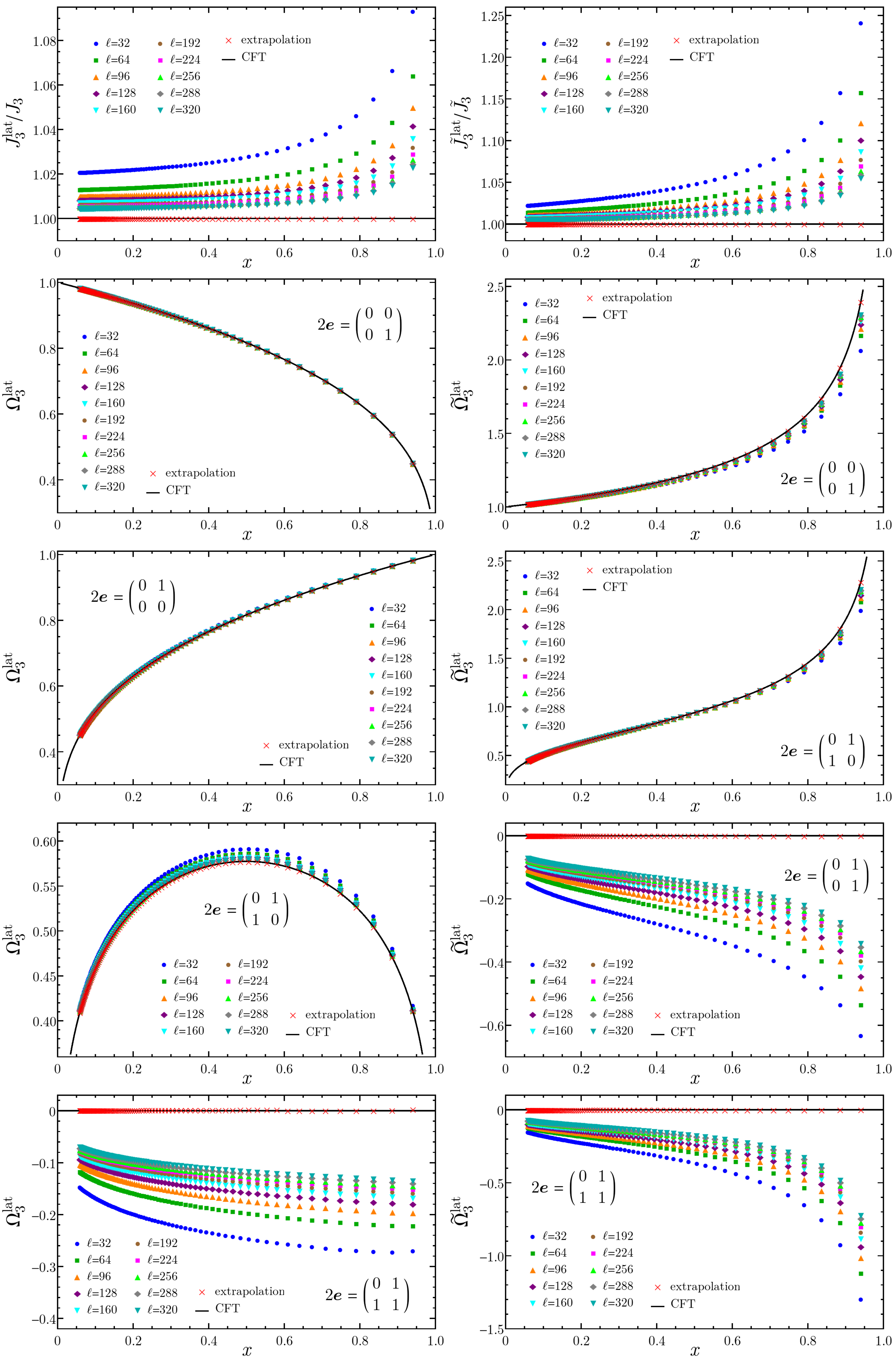}
\vspace{-.3cm}
\caption{
The terms in $\Tr \rho_A^3$ (left) and $\Tr\big(\rho_A^{T_2}\big)^3$ (right) for the critical Ising chain (see (\ref{eq:TrrhoA n3}) and (\ref{eq:TrrhoAT n3})  respectively).
For each group of identical terms, only one representative is shown.
Further details are given in the caption of Fig.\,\ref{fig:Is n2 all}.
} 
\label{fig:Is n3 all}
\end{figure}

The numerical results for the XX model are reported in Figs.\,\ref{fig:XX n2 all} and \ref{fig:XX n3 all} for $n=2$ and $n=3$ respectively;
while in Figs.\,\ref{fig:Is n2 all} and \ref{fig:Is n3 all} the same quantities are plotted for the Ising model.
As for the prefactor, the ratios $J_n^\text{lat}/J_n$ and $\tilde J_n^\text{lat}/\tilde J_n$ have been plotted  in order to eliminate the residual $\ell$ dependence. 
As for the remaining terms, only one representative for each equivalence class has been considered.
We have not shown the case $n=4$ because there are too many terms to consider. 
The analysis of the latter case, which is more complicated and not very illuminating, can be done by considering the explicit formulas of $\Tr\rho_A^4$ and $\Tr(\rho^{T_2}_A)^4$ for the lattice written explicitly in \cite{fc-10} and \cite{ctc-15} respectively.

The points computed on the lattice tend to the corresponding CFT predictions as the subsystem size increases.
This conclusion can be drawn only once the finite size effects have been taken into account through an accurate scaling analysis. 
Such analysis has been performed in \cite{fc-10} for $ \Tr \rho_A^n $, in \cite{ctc-15} for $\Tr ( \rho_A^{T_2})^n$  and in \cite{ctc-15-2} term by term in the formula found for the free Dirac fermion.
It is well established \cite{ccen-10,ce-10,un-vari} that the scaling in $\ell$ of these quantities is a power law governed by some unusual exponent $\delta_n=2h/n$.
In  \cite{cc-10} it has been found through general CFT arguments that such unusual corrections come from the insertions of the relevant operator with smallest scaling dimension $h$ at the branch points.

In the XX model the terms of the form $\ell^{-2m/n}$ occur for any positive integer $m$.
Since these terms converge slower and slower as $n$ increases, we have to include many of them in the scaling function.
The most general  ansatz for $\widetilde \Omega_n^\text{lat}$ of the XX model at finite $\ell$ in \eqref{J def lattice XX} takes the following form
\begin{equation} 
\label{eq:scaling Omega XX}
\widetilde\Omega_n^\text{lat}[\tilde{\boldsymbol q}]^2 
= \widetilde\Omega^2_n[2 \boldsymbol e] + \frac{\omega^{(1)}_n(x)}{\ell^{2/n}} + \frac{\omega^{(2)}_n(x)}{\ell^{4/n}} + \frac{\omega^{(3)}_n(x)}{\ell^{6/n}} + \dots \, ,
\end{equation}
where $\tilde{\boldsymbol q}$ and $\boldsymbol e$ are related as explained in \S\ref{sec:moments pt2}.
A scaling function similar to \eqref{eq:scaling Omega XX} can be studied also for the term $\tilde J_n^\text{lat}/\tilde J_n$.
Expressions analogous to (\ref{eq:scaling Omega XX}) can be defined for the terms in the formula of the moments $\Tr \rho_A^n $ of the reduced density matrix.
Fitting the data with \eqref{eq:scaling Omega XX}, the more terms we include, the more precise the fit could be.
Nevertheless, since we have access only to limited values of $\ell$, by employing too many terms overfitting problems may be encountered, leading to very unstable results. 
The number of terms to include in \eqref{eq:scaling Omega XX} has been chosen in order to get stable fits.
We find that every term $\widetilde\Omega^2_n[\boldsymbol e]$ follows the scaling \eqref{eq:scaling Omega XX} and the extrapolated value agrees with the corresponding CFT prediction.

As for the Ising model, also the Majorana fermion operator with conformal dimension $h=1/2$ must be taken into account  \cite{fc-10}.
This leads to a modification of the scaling function which includes more severe contributions of the form $\ell^{-m/n}$.
Thus, the most general  ansatz for $\widetilde \Omega_n^\text{lat}$ of the Ising model at finite $\ell$ in \eqref{J def lattice ising} reads 
\begin{equation} 
\label{eq:scaling Omega Ising}
 \widetilde\Omega_n^\text{lat}[\boldsymbol{\tilde q}] = \widetilde\Omega_n[2 \boldsymbol e] + \frac{\nu^{(1)}_n(x)}{\ell^{1/n}} + \frac{\nu^{(2)}_n(x)}{\ell^{2/n}} + \frac{\nu^{(3)}_n(x)}{\ell^{3/n}} + \dots\,,
\end{equation}
and analogously for $\tilde J_n^\text{lat}/\tilde J_n$ and the corresponding terms in the R\'enyi entropies.
Also in this case the agreement with the CFT predictions is very good.
From the analysis of the terms~\eqref{J def lattice ising} and~\eqref{J tilde def lattice ising}, we find a peculiar and unexpected feature.
It turns out that in the sum \eqref{renyi T_n} only the terms whose continuum limit corresponds to odd characteristics, and therefore vanishes identically, follow the scaling (\ref{eq:scaling Omega Ising}).
Instead, all the terms whose continuum limit corresponds to even characteristics follow a milder scaling, as can be qualitatively seen from the figures.
It turns out that the scaling of the latter terms is governed by the same exponents $\ell^{-2m/n}$ occurring for the XX model (see \eqref{eq:scaling Omega XX}).
This behaviour has been observed for both the terms in $\Tr\rho_A^n$ and the ones in $\Tr (\rho_A^{T_2} )^n$.

%%%%%%%%%%%%%%%%%%%%%%%%%%%%%%%%%%%%%%%%%%%%%%%%%%%%%%%%%%%%%%%%%%%%%%

\section{Conclusions}
\label{sec:conclusions}

We have considered the reduced density matrix of two disjoint intervals and its partial transpose with respect to one interval for the fermionic
systems given by the Ising model and the modular invariant Dirac fermion, which are the scaling limit of the critical Ising chain and of the critical XX spin chain respectively.

The CFT expressions of the moments $ \Tr\rho_A^n$ and $ \Tr ( \rho_A^{T_2})^n$ for these model are known for arbitrary order \cite{cct-11, ctt-13, a-13, ctt-14} and they are written as sums over 
all the possible combinations of boundary conditions (either periodic or antiperiodic) along the homology cycles of a canonical homology basis for the underlying Riemann surface (spin structures).
The moments of the reduced density matrix and of its partial transpose for the corresponding spin models on the lattice, whose scaling limit provide the conformal field theory formulas recalled in \S\ref{sec:cft rev}, are also written as sums of computable terms at a given order $n$ \cite{fc-10, ctc-15}.

In this manuscript, we have described a systematic method to get the scaling limit of any term occurring in the lattice expressions for $ \Tr\rho_A^n$ and $ \Tr ( \rho_A^{T_2})^n$ as the partition function of the fermionic model on the  underlying Riemann surface with a specific spin structure (see (\ref{corr renyi eps}) and (\ref{corr pt2 moments})).
Numerical checks have been performed to support our conclusions, where the proper finite size scaling corrections have been taken into account. 
Our analysis allows also to recover the moments of the reduced density matrix and of the moments of its partial transpose for the free fermion, which have been found in \cite{ch-05} and \cite{ctc-15-2} respectively.

\section*{Acknowledgments}

We are grateful to Maurizio Fagotti for collaboration at an early stage of this project. 
AC and ET acknowledge Kavli Institute for Theoretical Physics for warm hospitality, financial support and a stimulating environment during the program {\it Entanglement in Strongly-Correlated Quantum Matter}, where part of this work has been done.
For the same reasons, ET and PC are grateful to Galileo Galilei Institute for the program {\it Holographic Methods for Strongly Coupled Systems}. 
PC and ET have been supported by the ERC under Starting Grant  279391 EDEQS.

\begin{appendices}

\section{Recovering the free fermion}
\label{sec:free fermion}

In this appendix we discuss a procedure to recover $\Tr \rho_A^n$ and $\Tr (\rho_A^{T_2})^n $ for the free fermion from the corresponding expressions for the critical XX spin chain, whose scaling limit is the modular invariant Dirac fermion. 
This method has been first employed in  \cite{ctc-15} on the lattice and here we extend it to the scaling formulas.

The $c=1$ free Dirac fermion is the scaling limit of the tight binding model at half filling on the lattice, whose Hamiltonian reads
\begin{equation} 
\label{eq:Ham ff}
H=\frac12 \sum_{i=1}^L \Big[c^\dagger_i c_{i+1}+ c^\dagger_{i+1} c_{i} \Big] \, ,
\end{equation}
where periodic boundary conditions have been assumed. 
In the scaling limit, the CFT formula for $\Tr \rho_A^n$ is (\ref{Fn}) with $\mathcal{F}_n(x)=1$ identically \cite{ch-05}, while $\Tr (\rho_A^{T_2})^n $ is given by (\ref{tr rhoAn T2 cft}) where $\mathcal{G}_n(x)$ has been found in \cite{ctc-15-2} and we do not find useful to report it here.

Since the peculiar feature of the critical XX spin chain with respect to the tight binding model is the occurrence of the string $P_{B_1}$ of Majorana operators, 
our prescription to recover the moments for the latter model from the ones of the former one is to replace the operator $P_{B_1}$ with the identity operator ${\bf 1}$ in the lattice expressions of $\Tr \rho_A^n$ and $\Tr (\rho_A^{T_2})^n $ for the XX model, which have been found in \cite{fc-10} and \cite{ctc-15} respectively.

In the scaling regime, according to the discussions reported in \S\ref{sec:renyi} and \S\ref{sec:moments pt2}, replacing the string $P_{B_1}$ with the identity operator corresponds to remove all the operators $(-1)^{F_{B_1}}$,  which have been placed  on the Riemann surfaces $\mathcal{R}_n$ and $\widetilde{\mathcal R}_n$ to identify the scaling limit formulas of the various terms in (\ref{renyi T_n}) and (\ref{ptA2 tilde T_n}).

\subsection{Moments of the reduced density matrix}
\label{app dec renyi}

Let us consider the moments $\Tr \rho_A^n$ of the reduced density matrix for the XX spin chain.
Replacing the string $P_{B_1}$ with the identity operator ${\bf 1}$, from (\ref{fake rhoA})  we have that $\rho_A^{B_1}$ becomes  $\rho_A^{\,{\bf 1}}$ 
and therefore (\ref{rho 1234 def}) tells us that every $\rho_3$ and $\rho_4$ must be replaced by $\rho_1$ and $\rho_2$ respectively.
Performing these substitutions in (\ref{renyi rho plus}), we are left with $\Tr\rho_1^n$, as expected from \cite{pascazio-08}.

In the scaling limit, we have to consider $\mathcal{F}^{\textrm{Dirac}}_n(x)$ for the modular invariant Dirac fermion \cite{cct-11, ctt-14} reported in (\ref{Fn Dirac}).
Let us focus on a fixed spin structure characterised by the vectors $\boldsymbol\varepsilon$ and $\boldsymbol\delta$. 
Since the $a$ cycles only intersect $B_1$ and do not intersect $A_2$, removing all the operators $(-1)^{F_{B_1}}$ means that the boundary conditions along all the $a$ cycles are fixed to antiperiodic ones.
Being the vector $\boldsymbol\varepsilon$ determined by the boundary conditions along $a$ cycles (see (\ref{corr renyi eps})), we conclude that replacing $P_{B_1}$ with the identity  corresponds to replace $\boldsymbol\varepsilon$ with  the vector $\boldsymbol 0$ in the scaling limit formula.
As for the $b$ cycles, since they do not intersect $B_1$, the boundary conditions along them do not get modified and, consequently, the vector $\boldsymbol\delta$ is left untouched. 
By replacing every vector $\boldsymbol\varepsilon$ with $\boldsymbol 0$ in the sum occurring in (\ref{Fn Dirac}), it becomes
\begin{equation}
 \sum_{\boldsymbol\varepsilon,\boldsymbol\delta} (-1)^{4\boldsymbol\varepsilon\cdot\boldsymbol\delta}\,
 \Theta\bigg[\hspace{-.1cm} \begin{array}{c}
 \boldsymbol{0} \\  \boldsymbol{\delta}  
\end{array}\hspace{-.1cm} \bigg]
(\tau)^2 
=
2^{n-1}
 \sum_{\boldsymbol\delta} \delta_{\boldsymbol\delta , \boldsymbol{0}} \;
 \Theta\bigg[\hspace{-.1cm} \begin{array}{c}
 \boldsymbol{0} \\  \boldsymbol{\delta}  
\end{array}\hspace{-.1cm} \bigg]
(\tau)^2
=
2^{n-1}\,
 \Theta\bigg[\hspace{-.1cm} \begin{array}{c}
 \boldsymbol{0} \\  \boldsymbol{0}  
\end{array}\hspace{-.1cm} \bigg]
(\tau)^2 \,,
\end{equation}
Combining this result with (\ref{Fn Dirac}), one finds that $\mathcal{F}_n(x) =1$ identically, which is the expected expression for the free fermion found in \cite{ch-05}.

We find it worth remarking the following peculiar feature of the procedure described above. 
The terms in the sum (\ref{Fn Dirac}) with odd characteristics, which vanish identically,  get an even characteristic once $\boldsymbol\varepsilon$ is replaced by $\boldsymbol 0$, and therefore they become non vanishing after such replacement. 
Thus, the choice of introducing minus signs in front of the terms with odd characteristic in (\ref{Fn Dirac}) is crucial to get the correct result for the free fermion through the method described here.

\subsection{Moments of the partial transpose}

Considering the moments $ \Tr ( \rho_A^{T_2})^n $ of the partial transpose for the critical XX spin chain, the effect of replacing all the $P_{B_1}$'s with the identity operators has been discussed in \cite{ctc-15},
where it has been first observed that the matrix $\rho_A^{T_2}$ becomes the sum of two Gaussian matrices given in \cite{ez-15} for the free fermion and then its moments have been studied.

In the scaling limit, considering the canonical homology basis $\{\tilde{a}_r, \tilde{b}_r \,; \, 1\leqslant r\leqslant n-1\}$ introduced in \cite{ctt-14} and adopted in this manuscript for $\widetilde{\mathcal{R}}_n$ (see Fig.\,\ref{fig:R4} for $\tilde{a}_r$ and the right panel of Fig.\,\ref{fig:bcycles} for $\tilde{b}_r$), let us remark that both the $a$ cycles and the $b$ cycles of $\widetilde{\mathcal{R}}_n$ intersect $B_1$ on one sheet at least.
Similarly to  \S\ref{app dec renyi}, since all the cycles $\tilde{a}_r$ intersect $B_1$ and do not cross $A_2$, removing all the operators $(-1)^{F_{B_1}}$ means that all the $\boldsymbol\varepsilon$'s in (\ref{tilde Fn Dirac})  must be replaced by $\boldsymbol 0$. 
Nevertheless, differently from Sec.\,\ref{app dec renyi}, now the removal of the $(-1)^{F_{B_1}}$'s affects also the boundary conditions along the $b$ cycles of $\widetilde{\mathcal R}_n$ (see (\ref{corr pt2 moments})).
In particular,  the vector $\boldsymbol{\delta}$ in (\ref{tilde Fn Dirac}) should be replaced by  $\boldsymbol{\delta}'\equiv (\boldsymbol\delta - \mathcal P \cdot  \boldsymbol\varepsilon)\textrm{ mod $2$}$, where $\mathcal{P}$ is a $(n-1) \times (n-1)$ matrix with $1$'s on the main diagonal, $-1$'s on the lower diagonal and $0$'s elsewhere.
As further consistency check, we notice that, by setting $\tilde{p}_j^{B_1} =0$ into (\ref{corr pt2 moments}), we recover the corresponding formula written for the free fermion in Eq.\,(53) of \cite{ctc-15-2}.

By applying the above substitutions, the sum over characteristics in (\ref{tilde Fn Dirac}) becomes
\begin{equation}
\label{pt2 ff sub}
 \sum_{\boldsymbol\varepsilon,\boldsymbol\delta} (-1)^{4\boldsymbol\varepsilon\cdot\boldsymbol\delta}
 \Theta\bigg[\hspace{-.1cm} \begin{array}{c}
 \boldsymbol{0} \\  \boldsymbol{\delta'}  
\end{array}\hspace{-.1cm} \bigg]
(\tilde{\tau})^2 
=
 \sum_{\boldsymbol\varepsilon,\boldsymbol\delta'} (-1)^{4\boldsymbol\varepsilon\cdot\left(\boldsymbol\delta'+\mathcal P\cdot \boldsymbol\varepsilon\right)}\,
 \Theta\bigg[\hspace{-.1cm} \begin{array}{c}
 \boldsymbol{0} \\  \boldsymbol{\delta'}  
\end{array}\hspace{-.1cm} \bigg]
(\tilde{\tau})^2 = 
 \sum_{\boldsymbol\delta'} t_n(\boldsymbol\delta')\,
  \Theta\bigg[\hspace{-.1cm} \begin{array}{c}
 \boldsymbol{0} \\  \boldsymbol{\delta'}  
\end{array}\hspace{-.1cm} \bigg]
(\tilde{\tau})^2\,,
 \end{equation}
where in the first step we have just changed the summation variable and in the second one we have introduced
\begin{equation}
 t_n(\boldsymbol\delta) \equiv \sum_{\boldsymbol\varepsilon} (-1)^{4\left(\boldsymbol\varepsilon\cdot\boldsymbol\delta + \boldsymbol\varepsilon \cdot \mathcal{P} \cdot \boldsymbol\varepsilon\right)} \,,
 \end{equation}
 which is the same expression given in Eq.\,(60) of \cite{ctc-15-2}.
It is not difficult to observe that $\boldsymbol\varepsilon \cdot \mathcal{P} \cdot \boldsymbol\varepsilon = \boldsymbol\varepsilon \cdot \mathcal{Q}/2 \cdot \boldsymbol\varepsilon$, where $\mathcal{Q}$ has been defined below (\ref{eq:Q}). 
Plugging the last step of (\ref{pt2 ff sub}) into (\ref{tilde Fn Dirac}), it is straightforward to recover the result of \cite{ctc-15-2} for the free fermion. 
Let us stress that the introduction of the minus signs in front of the terms with odd characteristics in (\ref{tilde Fn Dirac}) is crucial to obtain the result for the free fermion.

\end{appendices}

\end{document}